# The Classical Nature of Thermal Conduction in Nanofluids

Jacob Eapen
*Department of Nuclear Engineering*
*North Carolina State University, Raleigh, NC 27695*
jacob.eapen@ncsu.edu

Roberto Rusconi and Roberto Piazza
*Dipartimento di Ingegneria Nucleare*
*Politecnico di Milano, 20133 Milano (Italy)*

Sidney Yip
*Department of Nuclear Science and Engineering*
*Massachusetts Institute of Technology, Cambridge, MA 02139*

**Abstract**

Several new mechanisms have been hypothesized in the recent years to characterize the thermal conduction behavior in nanofluids. In this paper, we show that a large set of nanofluid thermal conductivity data is enveloped by the well-known Hashin and Shtrikman (HS) mean-field bounds for inhomogeneous systems. The thermal conductivity in nanofluids, therefore, is largely dependent on whether the nanoparticles stays dispersed in the base fluid, form linear chain-like configurations, or assume an intermediate configuration. The experimental data, which is strikingly analogous to those in most solid composites and liquid mixtures, provides a strong evidence for the classical nature of thermal conduction in nanofluids.

## 1.0 Introduction

The initial promise of nanofluids as an advanced, nanoengineered coolant has been tempered in the recent years by a lack of consensus on its thermal conduction mechanism. The early experiments showed a fascinating increase (up to 40%) in the thermal conductivity with low nanoparticle volume fraction (<1%) [1-3]. While several experiments with well-dispersed nanoparticles have shown modest conductivity enhancements consistent with the Maxwell mean-field theory [4-10], more instances of substantially larger enhancements are also reported in recent years [11-23] lending a reasonable confidence to the experimentations. The variability in these results thus, highlights the difficulty of preparation and reproduction of consistent or standardized nanofluid samples.

For nanofluids, the thermal conductivity enhancements beyond the prediction of Maxwell's theory [24] are often regarded to be anomalous or unusual. In the Maxwell theory, the effective thermal conductivity is estimated by combining a set of well-dispersed, spheres into one equivalent sphere. Originally developed for assessing the electrical conductivity, the Maxwell theory for the thermal conductivity of nanofluid is given by

$$\frac{\kappa}{\kappa_f} = \frac{1+2\beta\phi}{1-\beta\phi} \qquad (1)$$

where $\phi$ is the nanoparticle volume fraction, $\beta = [\kappa]/(\kappa_p + 2\kappa_f)$ and $[\kappa] \equiv \kappa_p - \kappa_f$ is the difference between the thermal conductivities of the nanoparticle and the base fluid. If a finite temperature discontinuity exists at the nanoparticle-fluid interface, Maxwell model would still apply provided one makes the substitution $\kappa_f \to \kappa_f + \alpha\kappa_p$ (on the right-hand side), where $\alpha = 2R_b\kappa_f/d$, $R_b$ is the interfacial thermal resistance, and $d$ is the nanoparticle diameter [25, 26].

In addition to larger thermal conductivity, experiments have also revealed more disagreements with the Maxwell model. The thermal conductivity is shown to have an inverse dependence with nanoparticle size [11, 18, 27, 28] and a linearly increasing behavior with temperature [22, 29]. Interestingly, there appears to be a fundamental difference between the thermal conductance behavior of solid composites and nanofluids. In the former, smaller dispersed (or filler) particles, especially those in the nanometer size range, significantly reduce the matrix thermal conductivity. In some cases the thermal conductivity is reduced well-below that of the base medium [30] at all volume fractions while in others the enhancement is severely suppressed [31]. The solid composite behavior is easily explained through the interfacial thermal resistance, $R_b$, that has an inversely dependence on the particle diameter as shown before [25, 32, 33]. Thus, decreasing the filler particle size will reduce the effective thermal conductivity of the solid composites and *vice versa*. For the case of nanofluids the experimental data indicates that the thermal conductivity increases with decreasing nanoparticle size, a behavior which is clearly at odds with the Maxwell model [11, 18, 27, 28]. A few experiments also have also shown that the nanofluid thermal conductivity is not correlated in a simple manner to that of the nanoparticle as predicted by the Maxwell model [12, 20]. A limiting behavior at higher volume fractions is also observed in nanofluids which is qualitatively different from that in solid composites. While the thermal conductivity displays a quadratic or power law behavior at higher volume fractions for solid composites [34, 35], it is known to rise rapidly at lower volume fractions and then saturate for several nanofluids [12, 14, 20].

Several mechanisms have been recently proposed to account for the excess thermal conductivity beyond the Maxwell prediction. These include Brownian motion of the nanoparticles [36, 37], fluid convection at the microscales [38-44], liquid layering at the particle-fluid interface [45-50], nanoparticle shape [51, 52], cluster agglomeration [53, 54], or a combination of aforesaid mechanisms [55-59]. A disconcerting aspect of having several theories is compounded by the ability of each postulated mechanism to match the experimental data accurately.

In this paper, we start by asking an elementary question- Is there an anomalous enhancement in the thermal conductivity for nanofluids? By analyzing a large body of data including those from our experiments, we show that almost all the reported thermal conductivity data are enveloped by the well-known Hashin and Shtrikman (H-S) mean-field bounds for inhomogeneous materials [60]. This observation strongly indicates that the nanoparticles can exist in several configurations ranging from a well-dispersed mode to a linear chain-like arrangement. Indeed, a variety of experiments support the presence of both configurations through SEM (scanning electron microscope) pictures [1, 2, 6, 7, 11, 12, 14, 19, 20, 27]. A comparison to the conduction behavior of solid-solid composites and liquid-liquid mixtures reveal that the H-S bounds, to a large extent, have a universal applicability. While this is recognized for solid-composites it is not so well-known for liquid mixtures. We further show that the conduction behavior in binary solid composites differs from that of nanofluids in one important aspect which is its susceptibility to interfacial thermal resistance for nanometer sized filler particles. The apparent anomalous behavior of nanofluids is thus shown to emerge from a rather confined viewpoint of assuming well-dispersed nanoparticles; removing this restriction and allowing chain-like morphologies for the nanoparticles vastly improves the ability of mean-field theories to predict much higher thermal conductivities. We thus, affirm the classical nature of thermal conduction in nanofluids and show that it is consistent to that of binary solid composites or liquid mixtures.

## 2.0 The Theoretical Framework for Thermal Conduction in Nanofluids

For a single component material, thermal conduction is described uniquely with the Fourier constitutive law, $\dot{q}'' = -\kappa \nabla T$ where $\kappa$ is the thermal conductivity and $T$ is the temperature. In a mixture heat can flow from multiple gradients, in addition to the temperature gradient, such as those resulting from concentration and external fields. The theoretical framework is provided by the linear phenomenological theory which postulates that the fluxes are linear homogeneous functions of the corresponding gradients. While an intrinsic thermal conductivity exists for the nanofluids, the measured thermal conductivity can include effects of diffusion, chemical reactions and other external fields. Since diffusion, directly and indirectly, is considered to be a key mechanism for the nanofluid thermal conductivity in several new theories [36-44], the theoretical framework is elaborated here to make a quantitative assessment of diffusion and chemical reactions on nanofluid thermal conductivity. The formalism is well-known and in this paper, it is adopted from deGroot and Mazur [61].

This linearity of fluxes and gradients is expressed as

$$\mathbf{J} \equiv \mathbf{L}\hat{\mathbf{X}} = \sum_{k=1}^{n} L_{ik}\hat{X}_k \qquad (i=1,2,...,n) \qquad (2)$$

where $\mathbf{J}$ and $\hat{\mathbf{X}}$ are the generalized flux and gradient vectors respectively. $\mathbf{L}$ is a matrix containing the phenomenological coefficients. The cross coefficients $L_{ik}$ and $L_{ki}$ are equal, following the Onsager's reciprocity hypothesis. The heat flux in a multi-component system is not defined uniquely, as hence, the thermal conductivity. A commonly accepted definition follows from the second law of thermodynamics and for an $n$ component system, it is given by [61]

$$\mathbf{J}_q = \hat{\mathbf{J}}_q - \sum_{k=1}^{n} h^k \mathbf{J}^k \qquad (3)$$

where $\hat{\mathbf{J}}_q$ is the heat flux which is usually measured in an experiment, $\mathbf{J}_q$ is the reduced heat flux and $\mathbf{J}$ is the mass flux, and $h$ is the partial enthalpy for each species. The difference between the $\mathbf{J}_q$ and $\hat{\mathbf{J}}_q$ represents the heat transfer due to diffusion and hence, $\mathbf{J}_q$ is responsible for the intrinsic conduction in a multi-component system. For components ($s$, $l$), abbreviated for solid nanoparticles and base liquid, respectively, the phenomenological relationship reduces to the following form [61] since only temperature and concentration gradient are predominant in the nanofluid system

$$\mathbf{J}_q = -L_{qq}\frac{\nabla T}{T^2} - L_{qs}\frac{1}{T}\nabla(\mu^s - \mu^l)_T \quad (4)$$

$$\mathbf{J}^s = -L_{sq}\frac{\nabla T}{T^2} - L_{ss}\frac{1}{T}\nabla(\mu^s - \mu^l)_T \quad (5)$$

where $\mu$ is the chemical potential and for the binary nanofluid system, $\mathbf{J}^s = -\mathbf{J}^l$. In experiments, diffusion is associated with a concentration gradient ($c$) rather than the chemical potential. The above equations can be recast in terms of experimental coefficients in the following form [61]

$$\mathbf{J}_q = -\kappa\nabla T - \rho_s\left(\frac{\partial \mu^s}{\partial c^s}\right)_{T,p} TD''\nabla c^s \quad (6)$$

$$\mathbf{J}^s = -\rho c^s c^l D_T \nabla T - \rho D^{sl}\nabla c^s \quad (7)$$

The phenomenological coefficients are related to the experimental coefficients in the following way

$$\kappa = \frac{L_{qq}}{T^2} \quad (8)$$

$$D'' = \frac{L_{qs}}{\rho c^s c^l T^2} \quad (9)$$

$$D_T = \frac{L_{sq}}{\rho c^s c^l T^2} \quad (10)$$

$$D^{sl} = \frac{L_{ss}}{\rho c^l T}\left(\frac{\partial \mu^s}{\partial c^s}\right)_{T,p} \quad (11)$$

In Eq. (6) and 7, $D_T$, $D''$ and $D^{sl}$ stand for the thermal diffusion coefficient, the Dufour coefficient and the mutual (binary) diffusion coefficient, respectively, while the density of the system is given by $\rho$. Thermal diffusion coefficient ($D_T$) accounts for the flow of matter with a temperature gradient while the Dufour coefficient ($D''$) is a measure of the inverse effect which is the flow of heat due to the concentration gradient. The Onsager's reciprocity theorem ensures that the coefficients $D_T$ and $D''$ are equal. $\kappa$ is the thermal conductivity of the nanofluid system and it is clear that it is a native or intrinsic property to the nanofluid without any contribution from the diffusion. The ratio of $D_T$ to $D^{sl}$ is defined as the Soret coefficient and is given by

$$s_T \equiv \frac{D_T}{D^{sl}} = -\frac{1}{c^s c^l}\frac{\nabla c^s}{\nabla T} \quad (12)$$

In an experiment, measurements can be made at the beginning $(\nabla c^s = 0)$ or at the end $(\mathbf{J}^s = 0)$. Thus, multiple definitions for the effective thermal conductivities for the system can be defined for these limiting conditions which are as follows

$$\mathbf{J}_q = -\zeta\nabla T, \quad \hat{\mathbf{J}}_q = -\hat{\zeta}\nabla T \quad (\nabla c^s = 0) \quad (13)$$

$$\mathbf{J}_q = -\lambda\nabla T, \quad \hat{\mathbf{J}}_q = -\hat{\lambda}\nabla T \quad (\mathbf{J}^s = \mathbf{J}^l = 0) \quad (14)$$

The reduced heat flux $\mathbf{J}_q$ is associated with two thermal conductivities ($\zeta, \lambda$) and the heat flux, $\hat{\mathbf{J}}_q$ with $(\hat{\zeta}, \hat{\lambda})$. As mentioned before, the difference between the heat fluxes (Eq. 3) is solely due to the heat carried by the diffusing particles. It can be shown that [61]

$$\zeta = \kappa,$$
$$\hat{\zeta} = \kappa + D_T(h^s - h^l)\rho c^s c^l \quad (15)$$

$$\lambda = \kappa - \frac{(D_T)^2}{D^{sl}}\left(\frac{\partial \mu^s}{\partial c^s}\right)\rho(c^s)^2 c^l T \quad (16)$$
$$\hat{\lambda} = \lambda$$

Eqs. (15) and (16) correspond to the conditions, $\nabla c^s = 0$ and $\mathbf{J}^s = \mathbf{J}^l = 0$, respectively. Thus for a measurement when the mass fluxes are zero, the effective thermal conductivity ($\lambda$) is always less than the intrinsic thermal conductivity ($\kappa$). At the beginning of the experiment when there are no concentration gradients, the effective thermal conductivity ($\xi$), as expected, is equal to that of the intrinsic value. A steady-state experiment corresponds to $\mathbf{J}^s = \mathbf{J}^l = 0$, and theoretically, the measured thermal conductivity is given by $\lambda$ while a transient hot wire measurement is closer to the condition, $\nabla c^s = 0$.

Eqs. (15) and (16) show the effect of diffusion on the thermal conductivity of a nanofluid assuming that there are only two gradients, namely, temperature and concentration. The formalism is extendable to include other fields such as pressure gradient and external fields as long as the assumed linearity given in Eq. (2) is satisfied. In the next section, we will give an estimate of the various effective thermal conductivities for situations with simple diffusion and also with chemical reactions since nanosized particles are known to be highly reactive.

### 3.0 Possible Mechanisms for Nanofluid Thermal Conduction

#### 3.1 Mean-Field Models

The simplest and perhaps, the most intuitive mean-field models are the series and parallel modes of thermal conduction. In the former, the conducting paths, namely, those through the base fluid and the nanoparticles, are assumed to be in series and while in the latter they are regarded to be in parallel (see Fig. 1). The effective thermal conductivity of the nanofluid medium in the two models are given by [62]

$$\frac{1}{\kappa^{=}} = \frac{(1-\phi)}{\kappa_f} + \frac{\phi}{\kappa_p} \quad (17)$$

$$\kappa^{\parallel} = (1-\phi)\kappa_f + \phi\kappa_p \quad (18)$$

where $\kappa^{=}$ and $\kappa^{\parallel}$ are the series and parallel mode thermal conductivities, respectively.

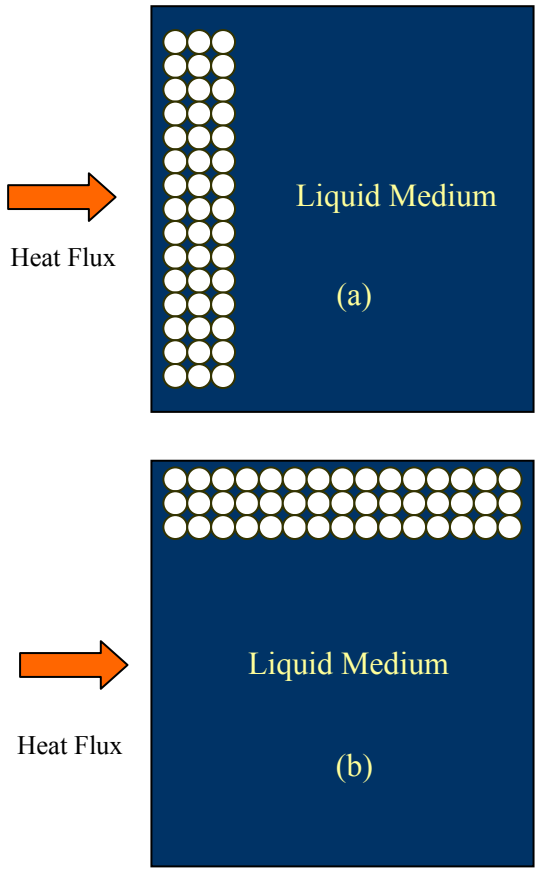

Fig. 1. A two-dimensional representation of series (a) and parallel (b) modes of conduction paths for nanofluids. The parallel mode represents the most efficient way of heat conduction in a binary nanofluid system.

As can be inferred from the SEM pictures, neither series nor parallel configuration is strictly applicable to a nanofluid even though the intertwined fibers allow the nanotube suspensions to be approximated by the parallel mode. Since the parallel mode corresponds the geometric configuration that allow the most efficient way of heat propagation in a binary system, it represents the absolute upper limit for the effective thermal conductivity regardless of the phase of the material. For example, the experiments in [63] show an increase of 5000% in the thermal diffusivity for low density polyethylene when the polymer fiber was configured in the direction of heat flow as opposed to perpendicular to the heat flow direction. Not surprisingly, numerous experiments have shown that the upper bound is rarely, if not ever, violated in binary solid composites or liquid mixtures.

The series and parallel bounds are not the narrowest that can be estimated with the mean-field approach. Hashin and Shtrikman (HS) has derived a set of bounds, that is most restrictive on the basis of volume fraction alone. Any improvement on these bounds would require additional knowledge on the statistical variations of the dispersed medium. The HS bounds for nanofluid thermal conductivity are given by [60]

$$\kappa_f\left(1 + \frac{3\phi[\kappa]}{3\kappa_f + (1-\phi)[\kappa]}\right) \leq \kappa \leq \left(1 - \frac{3(1-\phi)[\kappa]}{3\kappa_p - \phi[\kappa]}\right)\kappa_p \quad (19)$$

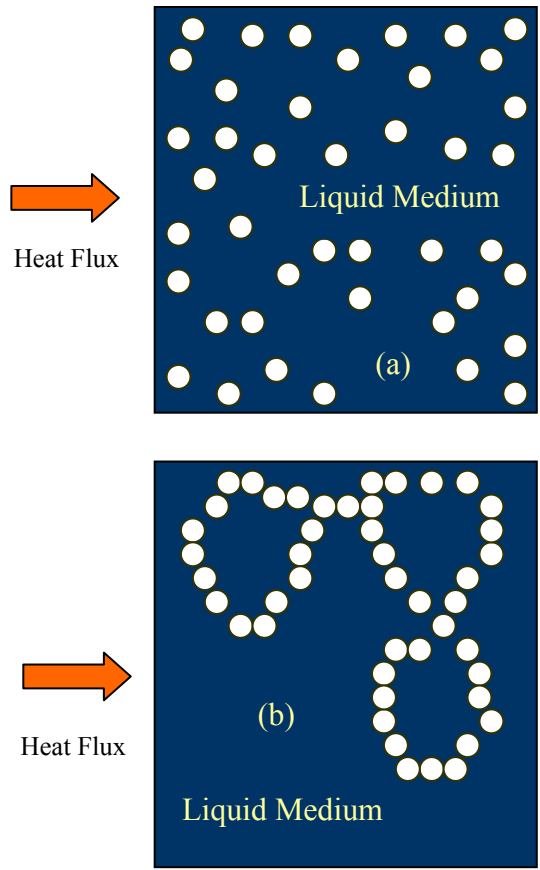

Fig. 2. A two-dimensional representation of the nanofluid configuration for the lower HS bound (a) and upper H-S bound (b). The lower HS bound corresponds to the Maxwell model when $\kappa_p > \kappa_f$, and *vice-versa*.

It is assumed that $\kappa_p > \kappa_f$ or otherwise, the upper and lower bounds would simply reverse. Notice that Eq.(1), which coincides with the lower HS bound when $\kappa_p > \kappa_f$ and with the upper bound in the opposite case, is rigorously exact to first order in $\phi$, as evident from the dilute limit, $\kappa = \kappa_f (1 + 3\beta\phi)$. Physically, the lower limit corresponds to a set of well-dispersed nanoparticles in a fluid matrix while the upper limit corresponds to large pockets of fluid separated by linked or chain-like nanoparticles as shown in Fig. 2.

In the lower HS configuration, the nanoparticles are always well-dispersed and therefore, the effective conductivity is biased towards the conduction paths in the surrounding fluid [64]. Like wise, the upper HS conductivity is biased towards the conduction paths along the dispersed nanoparticles. The lower HS limit ($\kappa^{HS-}$) thus, lies closer to the series mode of conduction while the upper limit ($\kappa^{HS+}$) approaches the parallel mode. If the configuration is neutral, i.e., neither favoring the series nor the parallel mode, then the effective thermal conductivity ($\kappa^0$) would lie in between lower and upper HS bounds. This approach, attributed to Bruggeman and also sometimes known as the effective medium theory (EMT), predicts the thermal conductivity in the implicit form given in Eq. (20) [60]. Both the HS and the unbiased model do not have any restrictions on the volume fraction unlike that of Maxwell which is strictly applicable only to dilute nanofluid systems.

$$(1-\phi)\left(\frac{\kappa_f - \kappa}{\kappa_f + 2\kappa}\right) + \phi\left(\frac{\kappa_p - \kappa}{\kappa_p + 2\kappa}\right) = 0 \qquad (20)$$

In a nanofluid, the unbiased configuration would be a mix of well-dispersed nanoparticles and linear aggregation. All the mean-field models thus, correspond to different configurations of the dispersed medium and it can be shown for $\kappa_p > \kappa_f$ [60, 62]

$$\kappa^= < \kappa^{HS-} < \kappa^0 < \kappa^{HS+} < \kappa^{\parallel} \qquad (21)$$

The effective $\kappa$ for a well-dispersed nanofluid is not a function of $\kappa_p$ as given by the relation $\kappa_p = \kappa_f(1+3\phi)$ provided $\kappa_p / \kappa_f > 1$. However, $\kappa$ for aggregated nanofluids, as given by the upper HS bound, is a function of the nanoparticle thermal conductivity and a larger enhancement is theoretically predicted for $\kappa_p / \kappa_f > 1$.

Allowing for clustering effects, therefore, dramatically broaden the thermal conductivity range that is consistent with the mean-field approach. The recent model of Prasher *et al* [53] assumes a linear chain-like clustered configuration for the nanoparticles and is very similar to the upper HS configuration. The aggregation model, as expected, predicts a substantially higher thermal conductivity in comparison to Maxwell model. In the limit $(\phi\kappa_p / \kappa_f) > 1$, the aggregation model [53] predicts $\kappa / \kappa_f = (\phi/3)\kappa_p / \kappa_f$ which can be compared to $(2\phi/3)\kappa_p / \kappa_f$ and $\phi\kappa_p / \kappa_f$ for upper HS and parallel modes, respectively. Even though the aggregation model [53] agrees with numerical simulations with fractal clusters, the lack of direct comparison to experiments does not allow the model to probe the purported anomaly of nanofluid thermal conduction. Interfacial thermal resistance has not been taken into account in any of these models yet and if applicable, it is easily incorporated [25]. The interfacial resistance always reduces the effective thermal conductivity and hence, the bounds presented here are the highest for the appropriate configurations [65].

As discussed before, not all nanofluid samples are well-dispersed as evident from the data from SEM analysis which point to the existence of partial clustering and chain-like linear aggregation (Figs. 3 and 4). Thus there is a strong evidence (though not conclusive) that linear nanoparticle clustering can explain the thermal conductivity enhancement beyond that of Maxwell predictions.

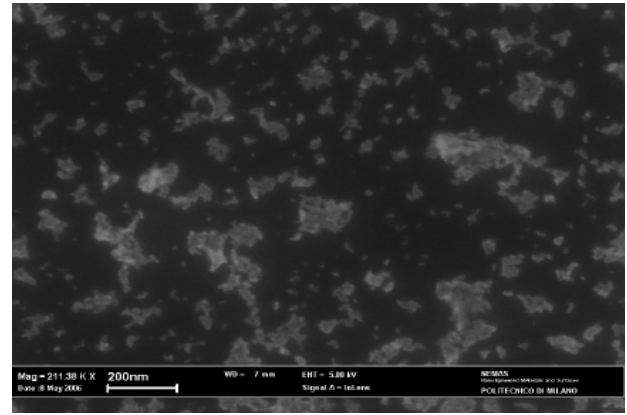

Fig. 3. SEM picture of alumina nanofluid

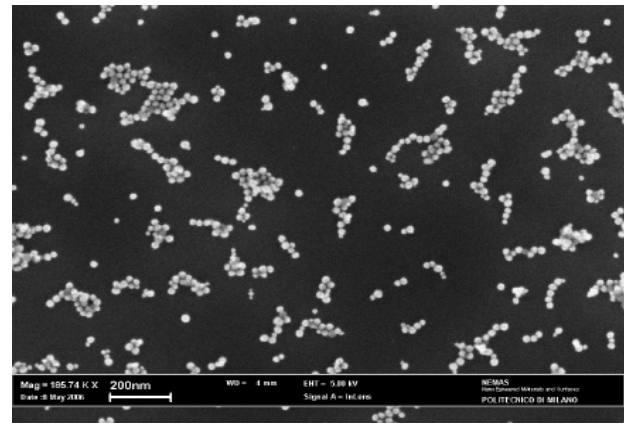

Fig. 4. SEM picture of Ludox (silica) nanofluid

In this paper, we will systematically explore the four mean-field bounds, namely, series, lower HS, upper HS and parallel, for the reported data on nanofluids and make a critical comparison with those observed in solid-solid composites and liquid mixtures. Analogous behavior will imply classical conduction mechanisms in all three binary systems while a persistent or conspicuous violations of the upper HS or parallel bounds, will give credence to the anomalous thermal conduction behavior in nanofluids.

**3.2   Interfacial Layer Models**

The interfacial layer models are a special case of mean-field models with the structure provided not by clustering of nanoparticles but through an ordered fluid structure around the nanoparticles [45-50, 55, 57, 58]. As such, the predictions of interfacial layer models are enveloped by the mean-field bounds discussed before. Interestingly, the interfacial layer models are far more popular than those assuming nanoparticle clustering (even though experiments suggest otherwise) and thus, are treated as a separate mechanism in this paper.

We will analyze the current state of knowledge to make a reasonable assessment on the role of interfacial layers in nanofluids. The motivation for proposing this model stems from both theoretical and experiment evidences of ordered layering near a solid surface. For example, molecular dynamics simulations predict three ordered layers of water on the Pt (111) surface [66]. In the first layer, water molecules form ice-like structures with the oxygen atoms bound to the Pt atoms while in the second and the third layers, water molecules that are hydrogen-bonded to the first and the second layer respectively, are observed. This ordering, even with a strong perturbation induced on the surface, persists over a distance of $O(1$ nm) from the Pt surface. Very similar behavior has been experimentally reported on a crystal-water interface [67]. Close to the interface, two layers of ice-like structures strongly bonded to the crystal surface were followed by two diffuse layers with less pronounced lateral and perpendicular ordering. Direct experimental evidences are also available for thin layering with liquid squalane [68] and non-polar liquids [69] adjacent to a solid surface.

In a recent molecular dynamics (MD) simulation it was shown that atomic-sized clusters of the order of 10 atoms, which interact strongly with the host fluid, produce an amorphous-like fluid structure [70] near the cluster. This interfacial structure, which is $O(1nm)$, provides a network of percolating conduction paths [70]. The effective thermal conductivity is much higher than that of the Maxwell prediction (50% enhancement at approximately $\phi$ =5%) but is well-bounded by the upper H-S limit. However, when the cluster size increases to hundreds of atoms, the percolation of interfacial structures is impeded dramatically and the relatively large enhancements reduce to more modest values consistent with Maxwell theory (lower HS bound). These results are in agreement with an earlier MD study [71], which showed no discernible increase in the interface thermal conductivity even while observing four ordered fluid layers near a surface. Similar results are also reported for water which showed no dependence of the molecular orientation on the thermal conductivity [72].

A theoretical estimate for the interface thickness around a nanoparticle can be made from the Yan model [58] which is given by

$$h = \frac{1}{\sqrt{3}} \left( \frac{4M_f}{\rho_f N_a} \right)^{1/3} \quad (22)$$

where $M_f$ and $\rho_f$ is the molecular weight and density of the surrounding fluid medium around a nanocluster and $N_a$ is Avagardo's Number given by $6.023 \times 10^{23}$. For water-based nanofluids the Yan model gives an interfacial thinkness of 2.84 nm which is larger that seen in experiments or MD simulations, but consistent on the order of magnitude.

Almost all the interfacial layer models which have been proposed in the past share the idea of a complex nanoparticle which is made of the bare nanoparticle and a postulated interfacial fluid nanolayer with an arbitrary thickness and thermal conductivity. The introduction of the nanolayer allows the estimation of an effective volume fraction which then is used in the regular mean-field models. Most models focus nanoparticles that are of the order 10 nm in size or smaller, and assume an interface thickness from 2-5 nm to provide evidence for a nanolayer-influenced thermal conduction mechanism. Sometimes, to make contact with the experiments an unsupported nanolayer thermal conductivity, as large as 10 times the fluid thermal conductivity [45] is also needed. However, the assumptions are at variance with the theoretical and experimental evidences [67-69, 73] which indicate that the interfacial layers are limited to a few molecular dimensions.

It is easy to show that an $O(1)$ nm layering is of no consequence to nanofluids that have been experimentally tested as the nanoparticles sizes are mostly of the order of 10 nm and above [74]. Since the volume fraction of the nanoparticles scales as $d^3$, a 1 nm interfacial thickness for a 10 nm nanoparticle would correspond to a relative volume change of $10^{-3}$ which is too small to account for any effect on the effective thermal conductivity. The experiments by Putnam *et al* [9] shows no unexpected increase in thermal conductivity (largely in agreement with Maxwell prediction) for well-dispersed gold particles that are as small as 4 nm. Without making a case of a strong interfacial thermal

Page 6

resistance, which we will analyze in more detail in the next section, and a subsequent postulation of an enhanced interfacial conductance that matches exactly with the Maxwell prediction, these experiments provide a direct evidence to the absence of any ordered liquid structures that influences the thermal conduction behavior. Indeed, such fluid structures, if they occupy a sizable volume in the fluid with a discernible density increase, they would have been detected by DLS. The absence of such structures in our measurements as well as those reported to date indicates that interfacial fluid structures which influence the thermal conductivity do not exist for the commonly tested nanofluids.

### 3.3 Brownian and Micro-convection Models

Effect of diffusion of nanoparticles

In a completely new approach, the Brownian and micro-convection models attempt to rationalize the temperature and size dependency by postulating a diffusion dependent thermal conductivity, albeit, different from that given by Eqs. (15) and (16). The Brownian models [36, 59] assumes that that the nanofluid thermal conductivity is dependent on the self diffusion coefficient of the nanoparticle which is given by the well-known Stokes-Einstein relationship, $D = k_B T / (3\pi\mu d_p)$ where $\mu$ is the dynamic viscosity, $d_p$ is the nanoparticle diameter and $k_B$ is the Boltzmann constant. Even though the natural outcome of this model is to predict the correct trends for temperature and nanoparticle size, this effort is criticized by several researchers, especially for the large mean free path of the liquid molecules with magnitude of $O(1)$ cm [75, 76]. In a Brownian dynamics (BD) simulation [37], the effect of diffusion is quantitatively estimated without resorting to explicit modeling. The simulations however, employs an algorithm that does not satisfy momentum or energy (the only quantity that is conserved is mass, and hence, the sole transport property that can be estimated from a BD simulation is the diffusion coefficient), and also employs a linear response (Green-Kubo) relationship for thermal conductivity that is not valid [77] for a BD phase space that is both discontinuous and stochastic [78].

As discussed in Section 2, diffusion can enhance the thermal conductivity due to Soret effect and also through chemical reactions that can occur between the nanoparticles and the solvent. In the simple diffusion case, the excess thermal conductivity due to diffusion is $\Delta\kappa = D_T (h^s - h^l)\rho c^s c^l$. Typical values for the colloids and nanofluid Soret coefficients are less than 0.1 K$^{-1}$ [9, 79, 80]. This gives an upper bound on the thermal diffusion coefficient $D_T$ of $O(10^{-10})$ m$^2$s$^{-1}$K$^{-1}$ with a diffusion coefficient $D^{sl}$ of $O(10^{-11})$ m$^2$/s corresponding to 10 nm sized nanoparticles. Note that the mutual and the self diffusion coefficients are nearly the same for small volume fractions. Since the difference in the specific enthalpy ($h$) between the nanoparticles is at most $10^4$ J/kg, the excess thermal conductivity $\Delta\kappa$ is several orders less than that of the base fluid.

At nanoscales, the nanoparticles can be exothermally reactive and diffusion accompanied by chemical reactions can also enhance the thermal conductivity [61]. If chemical equilibrium is reached before any appreciable transport, the effective thermal conductivity can increase by [61]

$$\Delta\kappa = \frac{\rho_f D^{sl} (\Delta h)^2}{T \left(\frac{\partial \mu^s}{\partial c^s}\right)} \quad (23)$$

where $\Delta h$ is the reaction heat which is of the order of the gradient in the chemical potential $\mu^s$ for the nanoparticles [61]. With a bounding $\Delta h$ of $10^5$ J/kg for reaction rates for nanoparticles in solutions (for example, the adsorption energy of water at alumina surface is ~140 kJ/mol [81]), $\Delta\kappa$ is again several orders less than that of the base fluid. Thus, there are sufficient theoretical reasons, as also noted from kinetic theory arguments [75, 82, 83], to believe that the direct diffusional motion of the nanoparticles has a negligible effect on the nanofluid thermal conductivity.

Effect of Micro-convection

In a related, but more intriguing hypothesis, the thermal conductivity is regarded to increase because of the large amount of the fluid that is carried by the diffusing nanoparticles [38-44]. In the 'micro-convection' hypothesis, it is hypothesized that convection currents set up by the Brownian motion of the nanoparticles can enhance the heat transfer between the nanoparticles and the base fluid, and hence, the nanofluid thermal conductivity. In this paper, we will focus on two micro-convection models that have received recognition in recent years. In the Jang and Choi model [39, 40], a new, but somewhat heuristic, heat transfer correlation is introduced to account for the randomly moving nanoparticles due to thermal motion which is given by

$$Nu \equiv \frac{hd}{\kappa_f} = O(\text{Re}^2 \text{Pr}^2) \quad (24)$$

The effective thermal conductivity can then be written as [40]

$$\frac{\kappa}{\kappa_f} = (1-\phi) + \frac{1}{(1+\hat{\alpha})}\left(\frac{\kappa_p}{\kappa_f}\right)\phi + C\left(\frac{d_f}{d_p}\right)\text{Re}_d^2 \text{Pr} \phi \quad (25)$$

where $\hat{\alpha} \equiv R_b \kappa_p / d$ stands for a non-dimensional interfacial thermal resistance, and $\text{Re}_d$ and Pr denote the Reynolds number for the nanoparticle and the Prandtl number for the base fluid, respectively. With negligible micro-convection and interfacial thermal resistance, Jang and Choi model coincides with the parallel mode of thermal conduction. A strong micro-convection effect will however, result in a thermal conductivity much higher than that of the parallel mode. Very good agreement is obtained with the experiments.

In the Prasher *et al* model [41, 42], the more traditional heat transfer correlation of flow over a sphere is adopted. Assuming that the Nusselt number on the scale of particle radius is $O(1)$, the Brownian motion of a single nanoparticle is regarded to increase the effective thermal conductivity of the base fluid by a factor of $[1+(1/4)\text{Re}\,\text{Pr}]$, A chief argument for micro-convection comes from the presumed presence of interfacial thermal resistance for the nanoparticles. Indeed, for nanosized filler particles in a solid composite, the effect of interfacial resistance can be very pronounced. Thus to account for the interfacial thermal resistance and the mixing of convection currents from multiple nanoparticles, the thermal conductivity of the nanofluid is fitted to experimental data using the expression [41]

$$\frac{\kappa}{\kappa_f} = \left(1 + A \text{Re}^\gamma \text{Pr}^{0.333} \phi\right)\left(\frac{1+2\beta\phi}{1-\beta\phi}\right) \qquad (26)$$

where $\gamma$ is a system-specific exponent, which for aqueous suspensions is found to have an optimal value of 2.5, and $A$ is constant attaining values as large as $4\times 10^4$. For negligible micro-convection effects and interfacial thermal resistance, the Prasher *et al* model reverts back to the Maxwell expression for well-dispersed nanoparticles.

We will now examine the characteristics of the micro-convection models in more detail. The hypothesized micro-convection effects appear through $\text{Re}_d = Vd/\nu$, where $V$ is the convection velocity and $\nu$ denote the kinematic viscosity of the base fluid. In both the micro-convection models, the convection velocities are represented by a 'Brownian velocity' to account for the rapidly oscillating nature of nanoparticle motion. In the Jang and Choi model, $V$ is given by [39]

$$V = \frac{D^{ss}}{l_f} = \frac{k_B T}{3\pi \mu d_p l_f} \qquad (27)$$

where $l_f$ is the mean-free path of a base fluid molecule and $D^{ss}$ is the self-diffusion coefficient of the nanoparticle. As noted by the authors, it is not clear whether this ratio actually represents a random velocity of the nanoparticle. In the Prasher *et al* model (as also in [38]), the conventional thermal velocity of the nanoparticle is taken as the convection velocity which is given by [41]

$$V = \sqrt{\frac{3k_B T}{m}} = \sqrt{\frac{18 k_B T}{\pi \rho d_p^3}} \qquad (28)$$

where $T$ is the temperature, $k_B$ is the Boltzmann constant, and $m$ and $\rho$ are the mass and density of nanoparticle respectively. Interestingly, a different form is adopted in the Patel micro-convection model [44] which is given by

$$V = \frac{2k_B T}{\pi \mu d_p^2} \qquad (29)$$

The latitude in choosing several forms, and therefore, differing physical characteristics, stems from the non-rigorous concept of a Brownian velocity. In the formal theory of Langevin dynamics, a fluctuating thermal velocity is uniquely defined while velocities constructed based on diffusion characteristics are not.

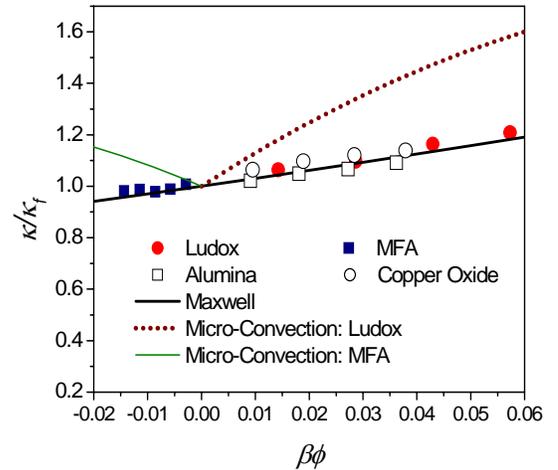

Fig. 5. THW data for Ludox ($\rho\sim 2200\,\text{kg/m}^3$, $d=32$ nm), MFA ($\rho\sim 2140\,\text{kg/m}^3$, $d=44$ nm), $Al_2O_3$ ($\rho\sim 4000\,\text{kg/m}^3$, $d=38$ nm) and CuO ($\rho\sim 6300\,\text{kg/m}^3$, $d=29$ nm) suspensions, plotted as a function of $\beta\phi$. The deviation of micro-convection model from Maxwell for $Al_2O_3$ and CuO are comparable to the experimental uncertainty (results not shown).

Two peculiar consequences of Eq. (28) are that for a given base fluid, temperature and nanoparticle size, the enhancement in the thermal conductivity increases with decreasing nanoparticle density $\rho$, and for nanoparticles with low density, the thermal conductivity can be largely positive even if $\kappa_p < \kappa_f$. In our previous paper [84], we had explicitly tested this prediction with transient hot-wire (THW) technique on nanofluids with silica and

Page 8

MFA (a copolymer of tetrafluoroethylene and perfluoromethylvinylether) nanoparticles that are lighter than the commonly tested alumina and copper oxide nanoparticles. We will report the main results briefly. Eq. (1) predicts $\kappa_p/\kappa_f$ to be a universal function of $\beta\phi$ while the micro-convection model does not. In Fig. 5, we plot two sets of data for Ludox and MFA as a function of $\beta\phi$ along with the reported experimental data in the literature for alumina and copper oxide which have higher densities as previously mentioned. Remarkably, all the experimental data collapse on to a single line predicted by the classical Maxwell theory without any interfacial thermal resistance and regardless of the nanoparticle density (or size). However, assuming micro-convection contributions lead to system-dependent predictions which are strongly conflicting with the experiments.

We have attributed the over-prediction of the micro-convection model from ascribing the nanoparticle thermal velocities as the convection velocities in place of the significantly lower thermophoretic drift velocities [84]. Am important observation we made was that a nanofluid under equilibrium conditions, will not support any convection regardless of Brownian motion of the nanoparticles. We note that the mixing of fluid currents around a nanoparticle, a concept which is borrowed from the macroscopic fluid flow, is not appropriate at microscales because the dragging of fluid around nanoparticles can occur at equilibrium conditions [85] in a nanofluid while macroscopic fluid flow is always under non-equilibrium conditions. To put it differently, the dragging of the bulk fluid in a nanofluid is thermally driven while at macroscopic scales, it is gradient driven. As explained in Section 2, an introduction of a thermal gradient, non-equilibrium coupling between mass and heat transport takes place due to the Soret effect. Thus, a colloidal particle acquires a thermophoretic drift velocity given by $u_T = D_T \nabla T$ [80]. For nanoparticles which are sufficiently larger than the molecular dimensions, the small Knudsen number makes the no-slip interface conditions a reasonable approximation [86]. Thus in typical thermal conduction experiments, the micro-convection velocities can only be of the order of the thermophoretic velocities.

Compared to the magnitude of the strongly fluctuating thermal speed, the thermophoretic velocities are insignificant in a nanofluid as they are characteristic of the collective motion of fluid motion around several diffusing nanoparticles. Our optical thermal lensing measurements have yielded a value of $D_T \sim 10^{-12}$ m$^2$s$^{-1}$K$^{-1}$ for both Ludox and MFA colloids, which in typical THW experimental time scales corresponds to thermophoretic velocities as low as 1 nm/s while the assumed convection velocities in Prasher *et al* and Jang and Choi models are O(1) m/s and O(0.1) m/s, respectively. Theoretical estimates of colloidal drift speeds are also in the range of O($10^{-8}$) m/s [79] which is consistent with our experimental values. To show that the thermophoretic velocities are insignificant even for very small nanoparticles, we have performed non-equilibrium molecular dynamics simulations (NEMD) on a model system. The details of the simulation method are given in [70]. Fig. 6 shows the relative magnitudes of the typical instantaneous nanoparticle velocity and the corresponding thermophoretic drift velocity. At steady state the magnitude of the thermophoretic drift velocity is two orders smaller than that of the RMS value, thus verifying its negligible magnitude.

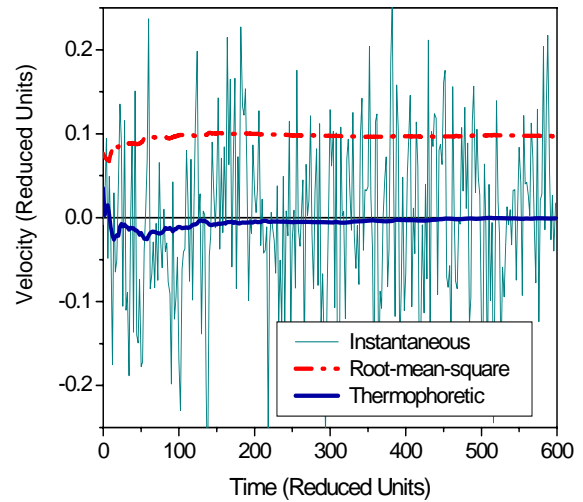

Fig. 6. Typical instantaneous and thermophoretic drift velocities in the *z*-direction of a 100 atom solid nanoparticle in a generic LJ fluid with NEMD simulations. The root-mean-square (RMS) velocity of the nanoparticle (0.098) is very close to $V = \sqrt{k_B T/m}$. With Argon fluid atoms, a reduced velocity of 0.1 and time of 100 corresponds to 15.79 m/s and 0.215 ns, respectively.

The fact that the particle thermal velocity is not the relevant velocity scale for heat transport is excluded by a simple argument. One should indeed compare the distance a particle move within a Brownian relaxation time $\tau = m/f$, where $f$ is the particle friction coefficient with the typical spatial scales of the thermal gradient (which are mesoscopic). For particles in the few tens of nanometer size range, $\tau \sim O(10^{-10})$ s, this time scale corresponds to a particle displacement of a few thousands of its diameter. The relaxation time $\tau$, which also sets the time decay of correlations between the particle momentum and energy density flux in the liquid [87], is negligible on the time scales probed by THW measurements. Thus, the difference of several orders of magnitude in the convection velocities precludes a significant contribution to the thermal conductivity from any conceivable micro-convection mechanism.

## 4.0 Mean-field Bounds for Nanofluid Thermal Conduction

In Section 3, we have provided experimental and theoretical evidences to show that the interfacial layers around a nanoparticle, Brownian motion of nanoparticles and micro-convection do not influence the thermal conductivity of the commonly tested nanofluids. In this section, we show that the mean-field theories are capable of explaining the rather large thermal conductivity enhancements reported so far. In addition, we also show that nanofluid thermal conduction behavior is strikingly similar to that observed in solid-composites and liquid mixtures by examining the theoretical bounds discussed in the previous section. However, we find that the interfacial thermal resistance, inferred from a large body of experimental data, is negligible for most nanofluids but can become significant for solid-composites.

## 4.1 Mean-field Bounds for Solid Composites

We will start by discussing the mean-field bounds for solid-composites which is a well-researched area for several decades. As discussed before the Hashin and Shtrikman (HS) limits are the narrowest bounds that can be constructed on the basis of volume fraction alone. The lower HS bounds is identically equivalent to the dilute Maxwell limit for well-dispersed nanofluids $(\kappa_p/\kappa_f \gg 1)$ while the upper HS bound corresponds to a linear chain-like configuration. The HS bounds, in turn, are enveloped by those from the series and parallel mode of conduction. The spread of the bounds depend on the relative thermal conductivities of the media. Most experiments with solid composites have large difference in the thermal conductivity and a relatively large spread, ranging several orders, can be expected for the enhancements.

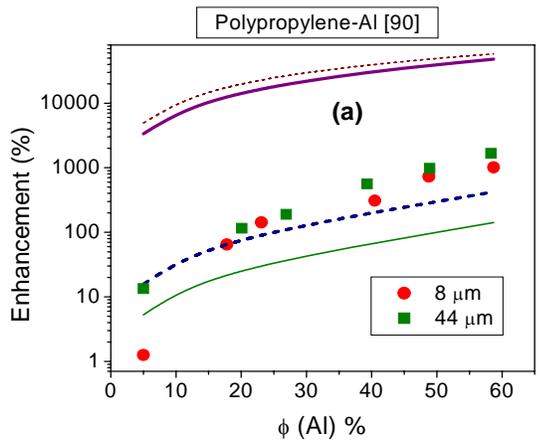

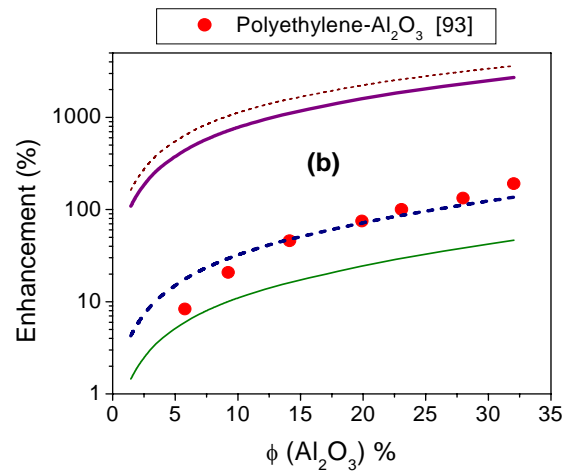

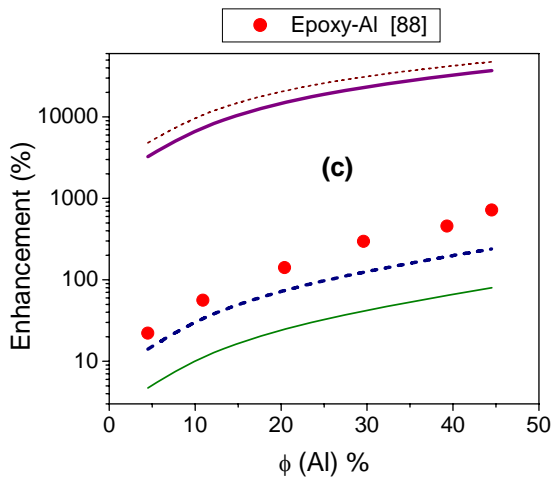

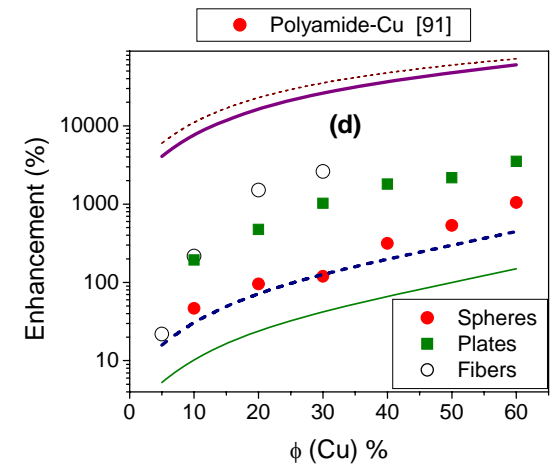

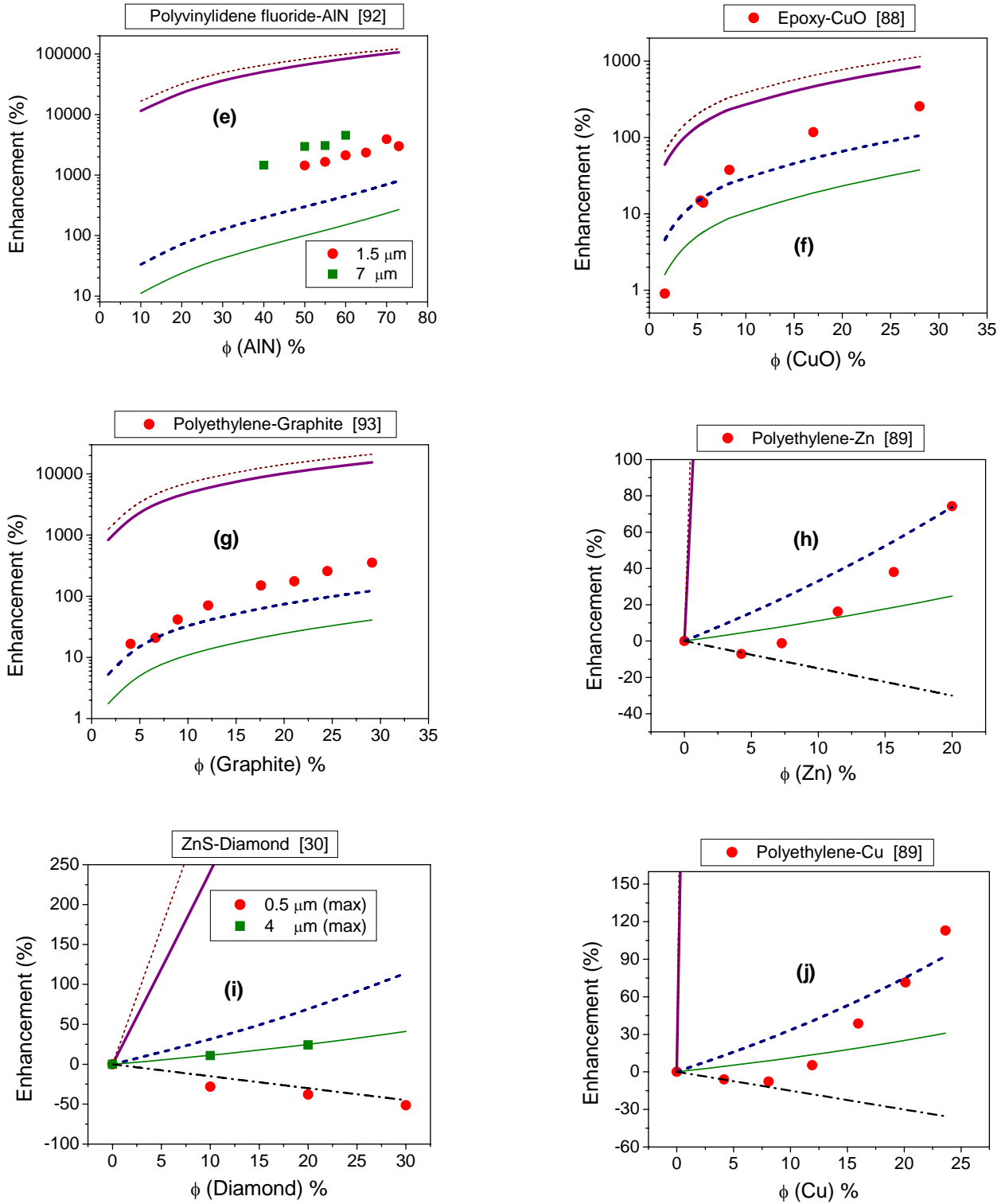

Fig. 7(a)-(j) Mean-field bounds for solid composites. The thin solid and thin dotted lines denote the enhancement in thermal conductivity with the series and parallel modes, respectively. The upper Hashin and Shtrikman (HS) bound is delineated by the thick solid line while the lower HS bound is given by thick dashed line. The experimental data [30, 88-93] are represented by the symbols. In (h), (i), (j), the lowest bound (dashed-dot line) is given by $(1-3\phi/2)$ for the limiting condition, $\alpha \to \infty$.

In Fig. 7, we have plotted the thermal conductivity enhancements for a large number of solid composite materials. The mean-field bounds are calculated based on the thermal conductivity data given in the Appendix. At larger volume fractions $(\phi > 10\%)$, the experimental data largely lie between the lower HS (Maxwell) and the upper HS bound. At lower $(\phi \leq 10\%)$, the thermal conductivity falls below the lower HS bound for a few composites, and for polyethylene-Cu, and polyethylene-Zn, and ZnS-Diamond, it becomes lower than the series mode prediction. Additionally, as the size increases, the thermal conductivity increases (Polypropalene-Al, Polypropalene-AlN, ZnS-Diamond) significantly.

All of the above observations can be reconciled within the frame-work of mean-field theories. The non-dimensional interfacial resistance parameter, $\alpha \equiv 2R_b\kappa_f/d$ determines the temperature discontinuity at the filler particle-base medium interface. In the limit $\kappa_p \gg \kappa_f$, the Maxwell expression becomes

$$\frac{\kappa}{\kappa_m} = \frac{(1+2\alpha) + 2\phi(1-\alpha)}{(1+2\alpha) - \phi(1-\alpha)} \qquad (30)$$

where $\kappa_m$ is the base medium thermal conductivity. Since $\alpha$ increases for decreasing filler particle size, $d$, the effective thermal conductivity decreases for smaller sized filler particles which is consistent with the experimental data on ZnS-Diamond and polyvinylidne fluoride-AlN. In the limit of $\alpha \to \infty$, $\kappa/\kappa_m$ reduces to $(1-3\phi/2)$. Thus, the effective thermal conductivity can become smaller than that of the base medium for all volume fractions as attested by data for ZnS-Diamond with 0.5 μm particles. Furthermore, the data for polyethylene-Cu and polyethylene-Zn are also bounded by $(1-3\phi/2)$.

Well-dispersed large spheres largely follow the lower HS limit except at higher volume fractions. As the volume fraction increases the filler particles tend to form chain-like configurations which promote a better heat transfer, and hence the thermal conductivity. This formation of interconnected or percolating filler particles explains the rapid non-linear increases in the thermal conductivity at higher volume fractions (see polyethylene-Cu, and polyethylene-Zn in linear scale). The filler materials in the form of fibers, as observed with Polyamide-Cu, are also very efficient for heat transfer. In a more dramatic observation, polyethylene polymer, a material having a low κ of 1 W/m-K, increases its thermal conductivity to 50 W/m-K, close to that of steel, when oriented in the direction of heat flow [63]. We emphasize here that the upper HS bound is never violated for any of the experimental data.

### 4.2 Mean-field Bounds for Liquid-Mixtures

Unlike those in solid-composites, the mean-field bounds in liquid-mixtures are not well-recognized. In the Sutherland-Wassiljewa theory, the thermal conductivity of a liquid mixture is given by [94]

$$\kappa = \kappa_1\left(\frac{n_1}{n_1 + A_{12}n_2}\right) + \kappa_2\left(\frac{n_2}{n_1 + A_{21}n_2}\right) \qquad (31)$$

where $n$ denotes the mole fraction and $1$ and $2$, stand for two components. The coefficient $A$, known as Wassiljewa coefficient, is given by the following semi-empirical form [95]

$$A_{ij} = \frac{1}{4}\left[1 + \left(\frac{\kappa_i}{\kappa_j}\right)^{1/2}\left(\frac{M_j}{M_i}\right)^{1/4}\right]^2\left[\frac{2M_j}{M_i + M_j}\right] \qquad (32)$$

where $M$ denotes the molecular weight. We can easily see that for equi-molar liquids and $A_{12}=A_{21}=1$, the above expression is identical to that of the parallel mode of thermal conduction.

In Fig. 8 we show the four bounds for a few representative liquid mixtures along with the unbiased estimate, Eq. (20). Most the data are nestled between the lower and higher HS bounds. Due to the small differences between the thermal conductivities of the two media, the bounds are relatively narrower. Liquids, in general, do not have identifiable structures even though X-ray and neutron scattering data show a structure for water possibly arising from a network of hydrogen-bonds. Such networked bonding can facilitate liquid molecules to form loosely formed dynamic structures with characteristics of both interconnected chains and isolated blocks of molecules. Such an arrangement will be consistent with that of the unbiased model. Indeed, we find that most of the data on liquid mixtures are well-predicted by the unbiased model which predicts a thermal conductivity that lies between the HS bounds.

The coefficient $A_{ij}$ in the Sutherland-Wassiljewa theory, originally developed for gaseous mixtures, has the physical interpretation of the ratio of efficiencies with which molecules $j$ and molecules $i$ impede the transport of heat [94]. The agreement with the unbiased estimate suggests that the coefficient $A_{ij}$ can be simply regarded as empirical factor in liquids to account for the dynamical structures.

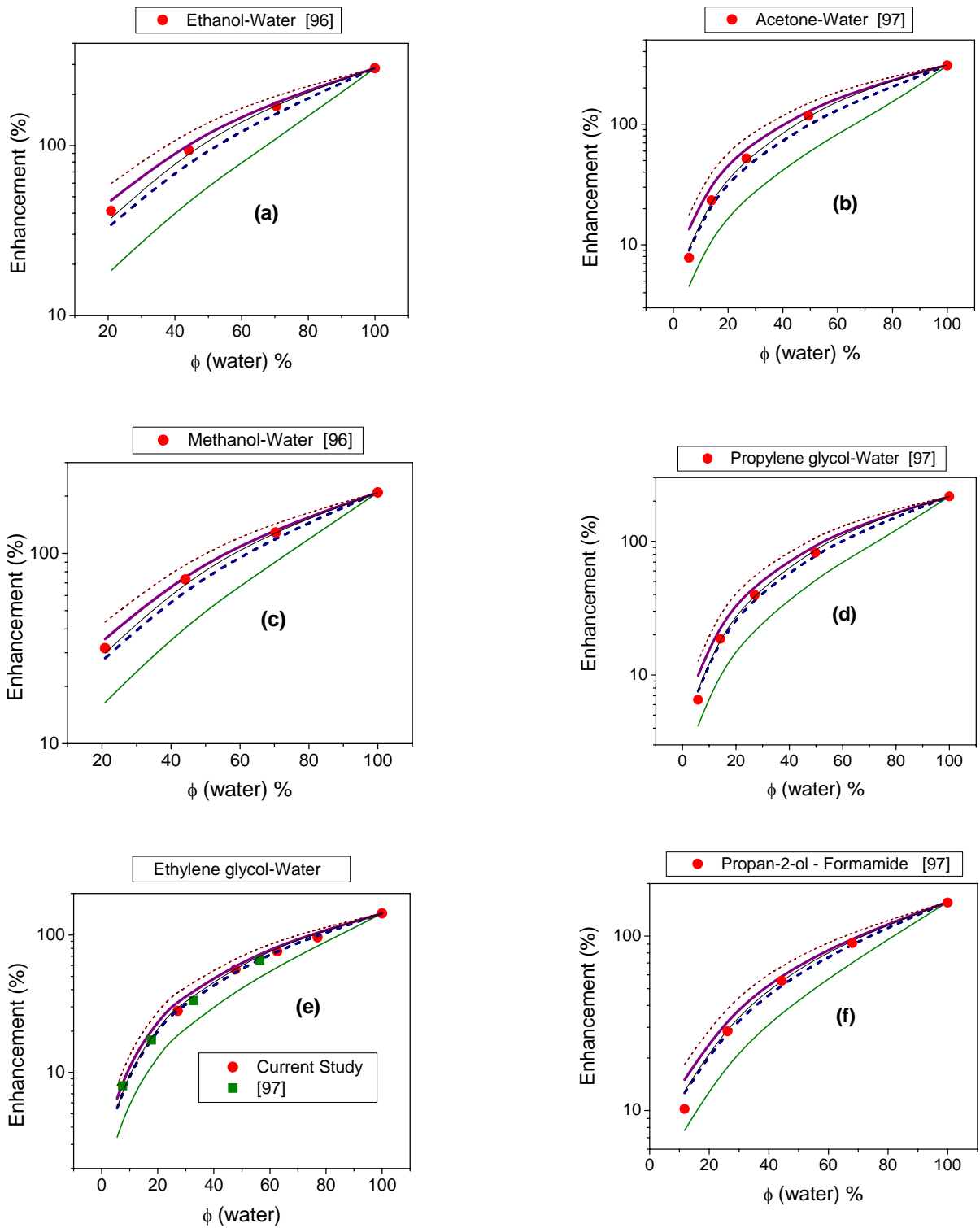

Fig. 8(a)-(f) Mean-field bounds for liquid mixtures. The line and symbol codes are the same as in Fig. 7 with the addition thin solid line between the HS bounds denoting the unbiased or Bruggeman estimate.
The experimental data are taken from [96] and [97].

### 4.3 Mean-field Bounds for Nanofluids

For both solid composites with negligible interfacial thermal resistance and liquid mixtures, HS bounds are respected to a large degree. However, these bounds have not been tested so far for nanofluids. In Fig. 9, we compare the mean-field bounds for a large body of nanofluid data, including those which have been described as unusual or anomalous. The data set includes oxide nanoparticles with relatively low $\kappa$ (silica, zirconia), moderate $\kappa$ (alumina, copper oxide), and high $\kappa$ (copper, aluminum, carbon nanotubes). It also includes different base media including water (polar), ethylene glycol and oil, and nanoparticles with lower thermal conductivity relative to the base media (C-60/70 and MFA in water).

Quite remarkably, all the data, except for a few sets, lie between the HS bounds affirming the same mechanism for thermal conduction for nanofluids as those of solid composites and liquid mixtures, namely, through molecular or electronic interactions. By examining the main features of the nanofluid data we can derive useful insights into the finer details of the conduction mechanism. The most striking feature is that only a small set of nanofluid data falls significantly below the Maxwell limit or lower HS bounds even at very low volume fractions and with nanoparticles that are in the tens of nanometers. This behavior is very unlike that in solid composites where at low volume fractions and nanometer sized filler particles, the effective thermal conductivity drop well below the series conductance bound. When the thermal conductivity of the dispersed media becomes closer to that of the base media, the HS bounds becomes narrower as can be noted with silica, zirconia and $Fe_3O_4$. This also implies that small errors (say 10-20%) in nanoparticle thermal conductivity can cause a discernible change in the mean-field bounds.

For well-dispersed nanoparticles, the enhancement is consistent with the Maxwell or lower HS bound. Since Maxwell limit represents the maximum thermal conductivity that that is possible with well-dispersed nanoparticles, it can be inferred that the interfacial thermal resistance for most nanofluids is negligible. Even if the nanoparticle thermal conductivity is smaller than that of the bulk value (as observed in nanosized thin films [98]), all the experimental data, except for fullerenes (C-60/70) and to some extent, MFA in water, remains bounded from below by the Maxwell theory with $R_b=0$. This observation is contrary to what has been presumed in the past, ostensibly from the strong influence of thermal resistance for solid composites and a rather limited nanofluid data set. Indeed, the original motivation of micro-convection hypothesis [41] stems from the presumed role of interfacial resistance in nanofluids.

Interfacial thermal resistance

The occurrence of an interfacial thermal (Kapitza) resistance at a liquid-solid interface has been experimentally evaluated by Cahill and co-workers who observed a bounding $R_b$ of $0.67\times10^{-8}$ $Km^2W^{-1}$ and $2\times10^{-8}$ $Km^2W^{-1}$ for hydrophilic and hydrophobic interfaces respectively [99]. With nanofluids with carbon nanotubes, a large variation in $R_b$, ranging from a low $0.24\times10^{-8}$ $Km^2W^{-1}$ [100] to a high $8.3\times10^{-8}$ $Km^2W^{-1}$ [101, 102] that is comparable to $R_b$ in a solid matrix (for example, diamond-silicon composite having an $R_b$ of $27\times10^{-8}$ $Km^2W^{-1}$ [98]) is also reported. The large span in the $R_b$ data, and the near zero $R_b$ inferred from Fig. 9, indicates that influence of the fluid interactions on the interfacial thermal resistance in nanofluids. This is further substantiated by the very different behavior for fullerenes (C-60/70) suspensions in surfactant- stabilized water, and oil without the use of a surfactant. While for fullerenes in water, a small, but identifiable reduction in thermal conductivity below the Maxwell prediction is recorded, possibly due to the interactions of surfactant molecules with surrounding liquid, fullerenes in oil is consistent with Maxwell prediction with $R_b=0$.

Theoretical studies show that $R_b$ attains relatively large values only when the liquid does not wet the solid surface. In our context, complete wetting may be a reasonable assumption for dispersions of hydrophylic colloids such as Ludox, and possibly for charged MFA colloids, where particle solvation is ensured by electrostatic forces. We also point out that terms like "hydrophobic" and "hydrophilic" are rather subtle, and the macroscopic concepts such as the contact angle may be a bit misleading. The rate of energy transfer would be indeed weaker if the liquid does not wet the solid, since in this case the liquid density in the interfacial layer would be depleted. Yet, from a microscopic point of view, what one may need to consider is the free energy of insertion of the particle in the fluid. For a stable, non aggregating colloidal dispersion, the latter is certainly negative (meaning, the particles are well-solvated). This means that even particles made of a hydrophobic material such as MFA can behave as "hydrophilic". The reason for this apparent paradox is related to the presence of the charged double-layer, which leads to the formation of a solvation layer made of hydrated counterions, hindering solvent depletion in the interfacial layer. More studies are however, needed for a microscopic understanding of the solid-fluid interfacial resistance in the presence of solvation shells.

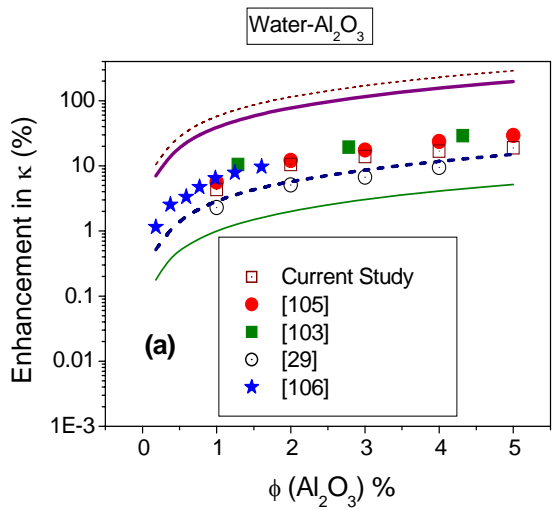
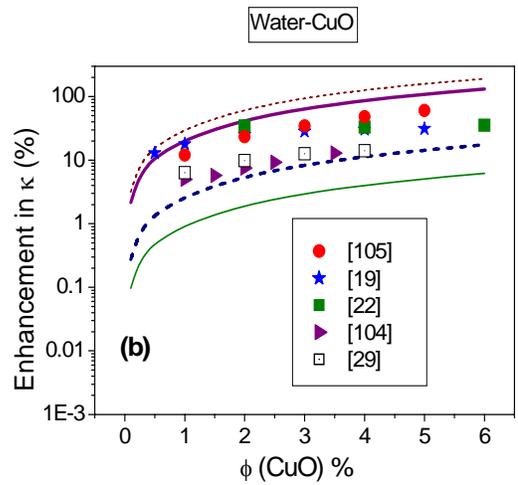
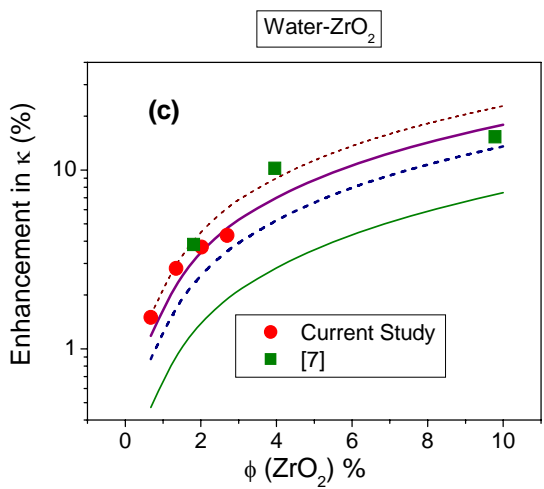
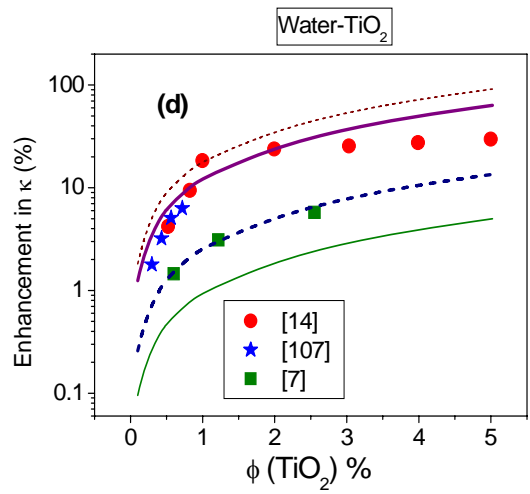
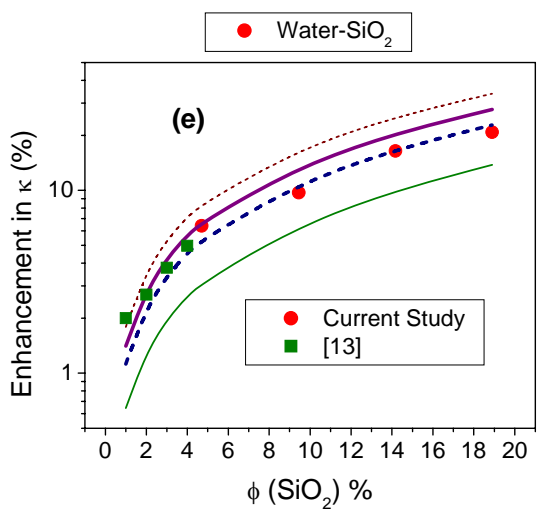
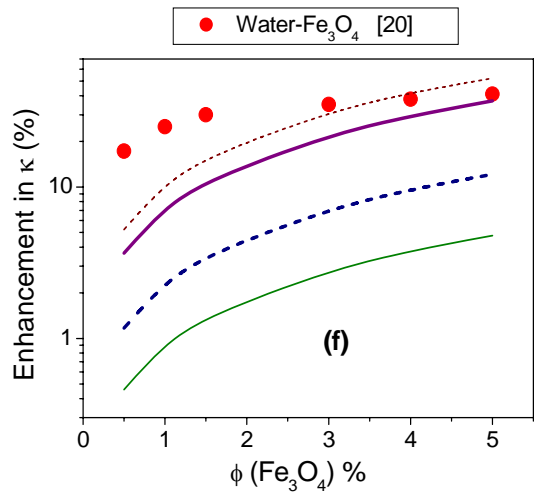

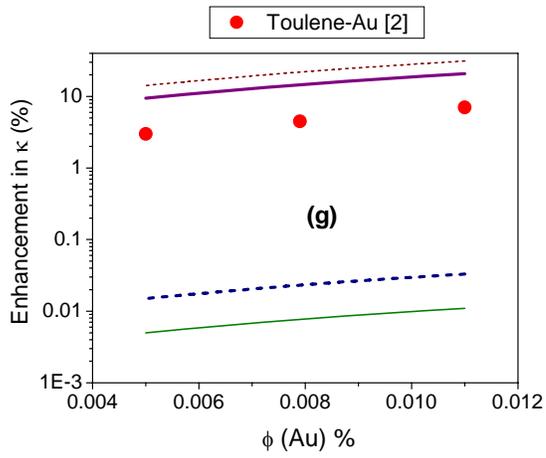
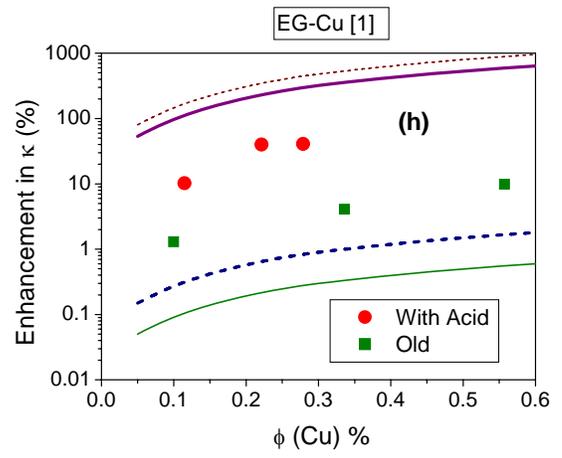
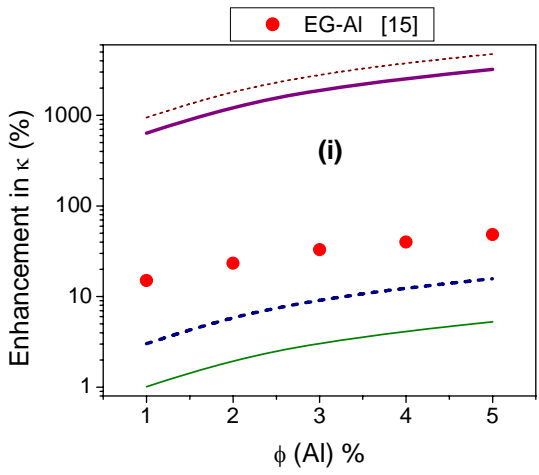
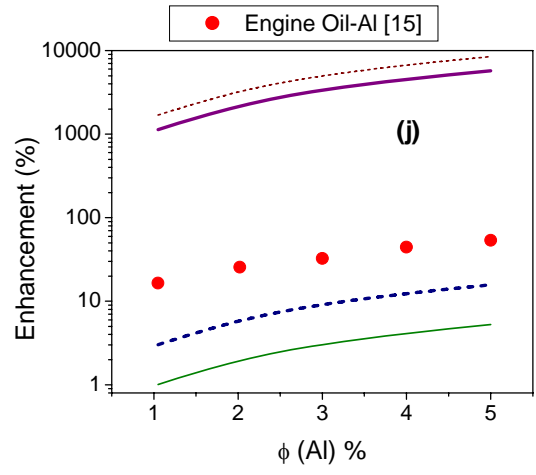
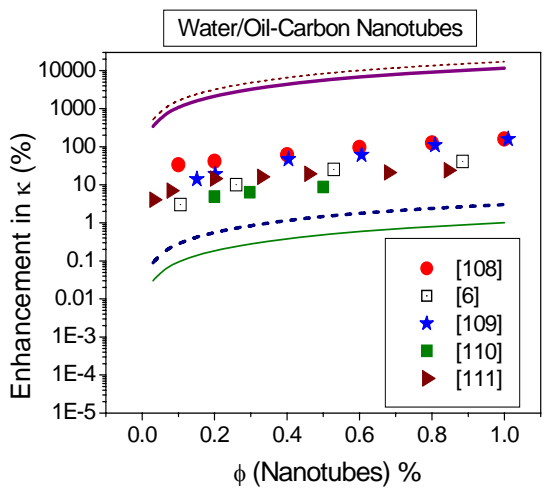
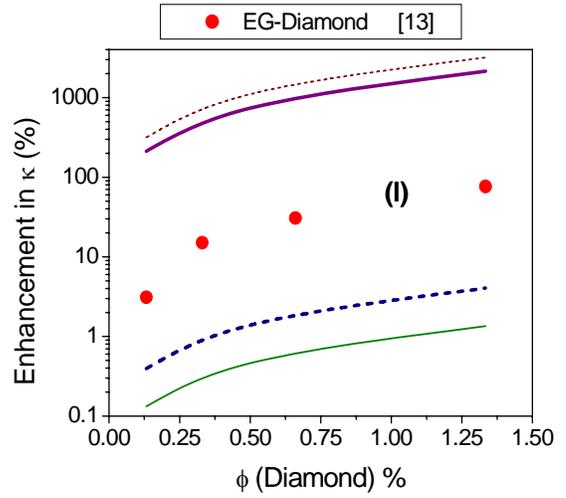

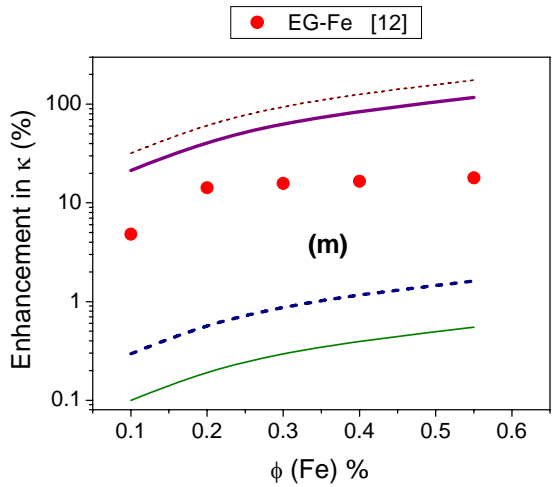
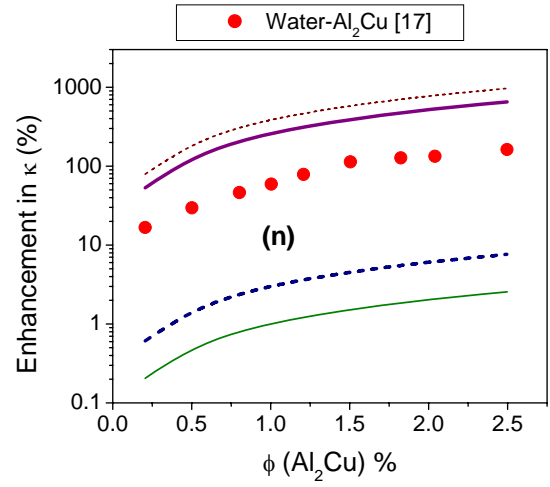
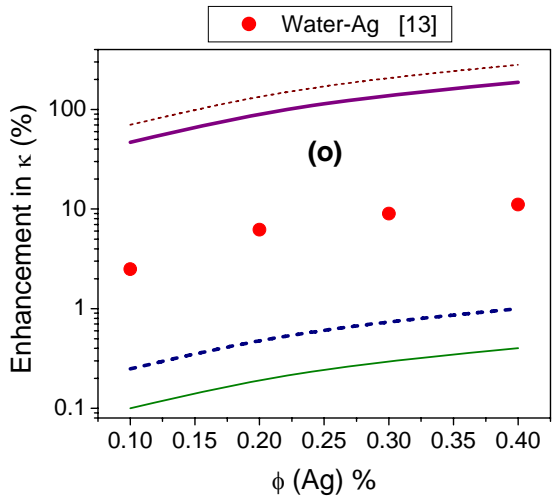
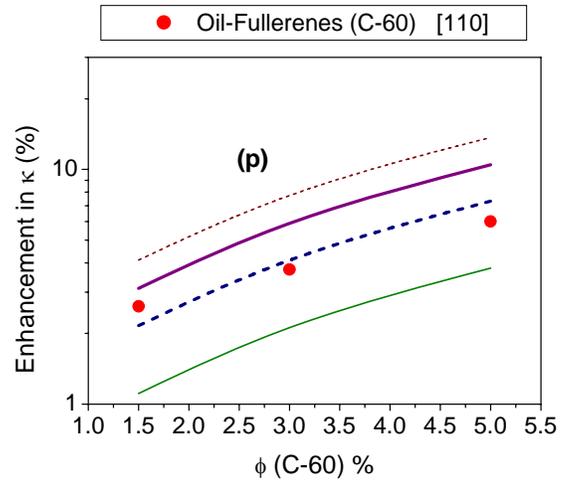
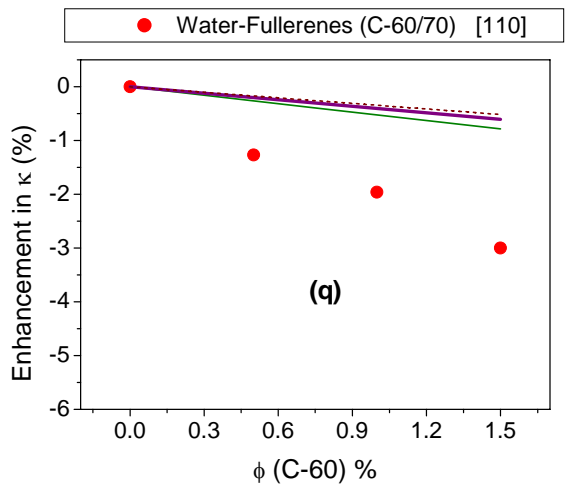
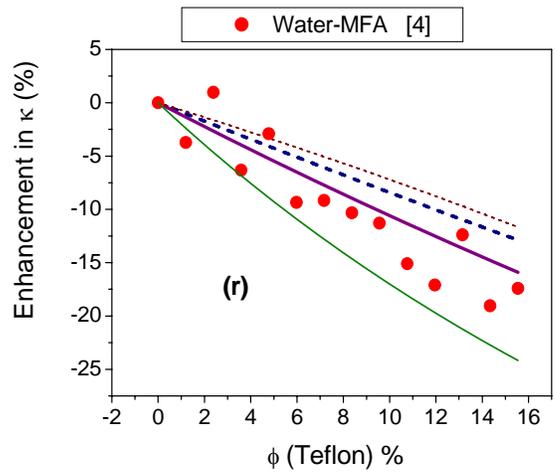

Fig. 9(a)-(r) Mean-field bounds for nanofluids. The line and symbol codes are the same as in Fig. 7. The references for experimental data are [19, 29, 103-106], [19, 22, 104], [7], [14, 107], [13], [20], [2], [1], [15], [6, 108-111], [12], [17] and [4].

A possible explanation for differences between the thermal resistance at a solid-solid and solid-liquid interface comes from the role of phonons (vibrational) modes with longer wavelengths in solids and its absence in liquids and dilute nanofluids. In classical terms, phonons are simply non-local, vibrational modes in a system. In the linear response theory [112], the thermal conductivity is the integral of autocorrelation of the microscopic heat flux vector which can be written as [113]

$$\lambda = \frac{1}{3Vk_BT^2}\int_0^\infty \langle \mathbf{J}_q(t)\mathbf{J}_q(0)\rangle dt \qquad (33)$$

where $\mathbf{J}_q$ is the microscopic heat flux vector, $V$ is the volume and $t$ is the time. According to Eq. (33), the longer the heat flux is correlated in time, measured by the product $\mathbf{J}_q(\tau)\mathbf{J}_q(0)$, higher will be the thermal conductivity. We will neglect the electronic contribution in the present discussion. The behavior of the heat flux autocorrelation (HACF) is not directly observable through experiments but is easily accessible from molecular dynamics simulations [113] provided accurate interatomic potentials are available. Our equilibrium MD simulations in Argon crystal, a material selected as a representative solid, the heat flux autocorrelation function portrays a two-stage behavior (see Fig. 7). The decay of the first stage is associated with the local dynamics [114] corresponding to the dynamics of nearest neighbors, while the slow decay of the second stage is associated with the collective dynamics of the long wave-length phonons that arise from the long-ranged crystalline order. In a liquid, there is no long ranged order and the heat flux autocorrelation shows a single stage, near exponential relaxation behavior as shown in Fig. 7. The details of the equilibrium MD simulations are given in [115].

For solid composites, the inclusion of filler materials acts as scattering centers and the contribution of long-wavelength phonons would be quickly impeded. Indeed, the introduction of even tiny amounts of impurities in a perfect lattice, impedes the long-ranged phonon contribution and dramatically reduce the thermal conductivity [116]. With a dilute nanofluid system, the HACF behavior it is reasonable to assume that the correlation function remains unaltered in the long-time behavior. Our MD simulations on a model nanofluid affirms this behavior by portraying a correlation behavior that is nearly exponential in nature, and a correlation time that remains more or less unchanged (see Fig. 11) with the inclusion of nanosized solid clusters. The cluster sizes in the MD simulation are of O(1nm) with a weak, but fully wetted cluster-fluid interaction and corresponds to a volume fraction of approximately 5%. The details of the model system are given in [70]. The quick decay of the HACF is also noted with a single, larger 2 nm nanoparticle $(\phi=10\%)$ [83], and in a Xe-Pt model nanofluid system [115].

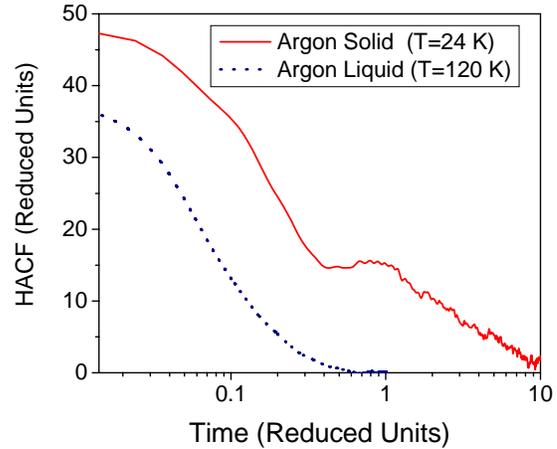

Fig. 10 Heat flux autocorrelation functions (HACF) for Argon solid and liquid from equilibrium MD simulations. The stronger correlation strength (measured by the magnitude) and longer time for perfect crystals assures a higher thermal conductivity, which is the area under HACF. A reduced time of 1 unit corresponds to 2.16 ps.

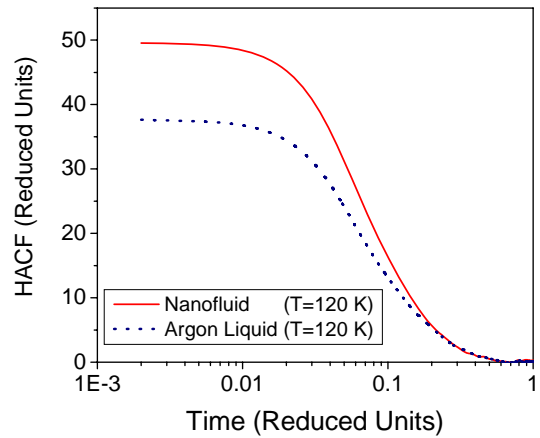

Fig. 11 Heat flux autocorrelation function (HACF) for a model nanofluid system from MD simulations. The near-identical correlation time indicates the absence of long-wavelength phonons with the insertion of solid clusters. A reduced time of 1 unit corresponds to 2.16 ps.

The strength of the HACF, in the present simulations increases due to a fully wetted cluster-fluid interaction resulting in an increase in the effective thermal conductivity consistent with the Maxwell prediction. For larger sized nanoparticles, an appreciable thermal resistance is noted only when the liquid does not wet the solid surface, as noted in a separate MD simulation [117]. The limited MD simulations, thus far indicates that the wetting properties of a solid-fluid interface plays a key role on the interfacial thermal resistance. Work is in progress for deriving a fundamental understanding of wetting and thermal resistance, especially within an electric double layer for charged colloids.

Resolving the Apparent Contradictions with the Mean-Field Approach

Coming back to the discussion on the experimental data on nanofluids, the apparent contradictions or anomalous behavior such as lack of correlation to particle thermal conductivity and size effects can be resolved by weighing in the ability of nanoparticles into forming linear chain-like clusters. Not all clusters are equally efficient in increasing the thermal conductivity as experiments show that the larger clusters without the linear chain-forming morphologies lead to a limiting behavior in the enhancement. The temperature dependence is also not as striking as it was earlier believed with the recent experiments [7, 118] showing a similar variation for both nanofluids and the base fluid. This implies that the mechanism for increase in the thermal conductivity of water (for example, the temperature dependent slow modes [119]) is also responsible for the thermal conductivity increase in the nanofluids as well. Conversely, it is reasonable to expect a decrease in the nanofluid thermal conductivity for a base fluid that has a negative change in thermal conductivity with increasing temperature. Additional experiments evidences are needed to verify this hypothesis.

## 5.0    Concluding Remarks

In conclusion, we have presented experimental evidence to show that almost all the reported nanofluid thermal conductivity data obeys the well-known Hashin and Shtrikman mean-field bounds for inhomogeneous materials. The striking similarity of the nanofluid data to solid composites and liquid mixtures strongly indicate that the mechanism of nanofluid thermal conductivity is classical in nature, namely, through molecular and electronic interactions. The earlier reports of anomalously high thermal conductivity can be traced back to an exclusive comparison of the test data to theories (such as Maxwell) that are applicable only to well-dispersed nanoparticles. Once this constraint is relaxed, and a linear chain-like configuration is allowed for the nanoparticles, the mean-field approach predicts a thermal conductivity range which easily accommodates almost all the reported data on nanofluid thermal conductivity.

A key difference between the thermal conduction behavior in a nanofluid and solid composites appears to be the interfacial thermal resistance. Again, a large body of nanofluid experimental data, including those from the current study, shows that the interfacial thermal resistance is negligible for most nanofluids while for nanosized fillers in solid composites, the effective thermal conductivity is strongly dependent on it. There are indications from molecular dynamics studies that the wetting properties have a significant influence on the interfacial thermal resistance in nanofluids. It is likely that hydrophilic nanofluids, such as those with oxide nanoparticles, and nanofluids that are electro-statically charged, are completely wetted due to presence of solvation layers. Molecular dynamics simulations are in progress to understand a fundamental understanding of interfacial thermal resistance in nanofluids, especially within an electric double layer.

It remains a challenge to accurately identify and manipulate the cluster configuration to modify the thermal transport properties of a nanofluid. The two characterization techniques, DLS and SEM have limitations in assessing the structure of nanoparticles. DLS measurements are limited to dilute suspensions ($\phi < 1\%$) for most nanofluids while SEM imaging can be performed only after drying the base fluid. On the basis of the experimental evidences provided in this paper, the key to having an enhanced thermal conductivity largely depends whether the nanoparticles are dispersed or form chain-like configurations. Within the current experimental techniques, it is possible to characterize the cluster formation through fractal dimensions [53, 59, 120]. Once these geometrical details, are available, a more precise comparison can be made between several samples that show appreciable differences in the thermal conductivity (for example, see the data on alumina and copper oxide nanofluids in Fig 9). While the science of making well-dispersed colloids have reached a fair amount of maturity, the attempts at generating targeted nanoparticle configurations in a nanofluid is still in an evolving phase. It is expected that future studies can systematically address the configurational constraints necessary for an enhanced thermal conduction in nanofluids.

**Acknowledgment**

J.E wishes to thank Jacopo Buongiorno, Wesley Williams, Pawel Keblinski and Ravi Prasher for insightful discussions on nanofluids. R.R acknowledges the generous support from the joint MIT/Politecnico "Progetto Rocca" funding.

# References

[1] J.A. Eastman, S.U.S. Choi, S. Li, W. Yu, L.J. Thompson, Anomalously increased effective thermal conductivities of ethylene glycol-based nanofluids containing copper nanoparticles, Appl. Phys. Lett. 78 (2001) 718-720.

[2] H.E. Patel, S.K. Das, T. Sundararajan, A.S. Nair, B. George, T. Pradeep, Thermal conductivities of naked and monolayer protected metal nanoparticle based nanofluids: Manifestation of anomalous enhancement and chemical effects, Appl. Phys. Lett. 83 (2003) 2931-2933.

[3] X. Wang, X. Xu, S.U.S. Choi, Thermal Conductivity of Nanoparticle-Fluid Mixture, J. of Thermophys. and Heat Transfer 13 (1999) 474-480.

[4] R. Rusconi, E. Rodari, R. Piazza, Optical measurements of the thermal properties of nanofluids Appl. Phys. Lett 89 (2006) 261916.

[5] D.C. Venerus, M.S. Kabadi, S. Lee, V. Perez-Luna, Study of thermal transport in nanoparticle suspensions using forced Rayleigh scattering, J. Appl. Phys. 100 (2006) 094310.

[6] X. Zhang, H. Gu, M. Fujii, Effective thermal conductivity and thermal diffusivity of nanofluids containing spherical and cylindrical nanoparticles, J. Appl. Phys 100 (2006) 044325.

[7] X. Zhang, H. Gu, M. Fujii, Experimental Study on the Effective Thermal Conductivity and Thermal Diffusivity of Nanofluids, Int. J. Thermophysics 27 (2006) 569-580.

[8] H. Xie, J. Wang, T. Xi, Y. Liu, Thermal conductivity of suspensions containing nanosized SiC partciles, Int. J. Thermophysics 23 (2002) 571-580.

[9] S.A. Putnam, D.G. Cahill, P.V. Braun, Z. Ge, R.G. Shimmin, Thermal conductivity of nanoparticle suspensions, J. Appl. Phys. 99 (2006) 084308.

[10] P. Keblinski, J.A. Eastman, D.G. Cahill, Nanofluids for thermal transport, Materials Today 8 (2005) 36-44.

[11] C.H. Chon, K.D. Kihm, S.P. Lee, S.U.S. Choi, Empirical correlation finding the role of temperature and particle size for nanofluid ($Al_2O_3$) thermal conductivity enhancement, Appl. Phys. Lett. 87 (2005) 153107.

[12] T.K. Hong, H.S. Yang, C.J. Choi, Study of the enhanced thermal conductivity of Fe nanofluids, J. Appl. Phys. 97 (2005) 064311.

[13] H.U. Kang, S.H. Kim, J.M. Oh, Estimation of thermal conductivity of nanofluid using experimental effective particle volume, Experimental Heat Transfer 19 (2006) 181-191.

[14] S.M.S. Murshed, K.C. Leong, C. Yang, Enhanced thermal conductivity of $TiO_2$-water based nanofluids Int. J. Therm. Sci. 44 (2005) 367-373.

[15] S.M.S. Murshed, K.C. Leong, C. Yang, Determination of the effective thermal diffusivity of nanofluids by the double hot-wire technique, J. Phys. D: Appl. Phys. 39 (2006) 5316-5322.

[16] M. Chopkar, P.K. Das, I. Manna, Synthesis and characterization of nanofluid for advanced heat transfer applications, Scripta Materialia 55 (2006) 549-552.

[17] M. Chopkar, S. Kumar, D.R. Bhandari, P.K. Das, I. Manna, Development and characterization of $Al_2Cu$ and $Ag_2Al$ nanoparticle dispersed water and ethylene glycol based nanofluid, Materials Science and Engineering: B 139 (2007) 141-148.

[18] C.H. Li, G.P. Peterson, The effect of particle size on the effective thermal conductivity of $Al_2O_3$-water nanofluids, J. Appl. Phys 101 (2007) 044312.

[19] H.T. Zhu, C.Y. Zhang, Y.M. Tang, J.X. Wang, Novel Synthesis and Thermal Conductivity of CuO Nanofluid, J. Phys. Chem. C 111 (2007) 1646-1650.

[20] H. Zhu, C. Zhang, S. Liu, Y. Tang, Y.Yin, Effects of nanoparticle clustering and alignment on thermal conductivities of $Fe_3O_4$ aqueous nanofluids, Appl. Phys. Lett. 89 (2006) 023123.

[21] Q. Li, Y. Xuan, Enhanced heat transfer behaviors of new heat carroer for spacecraft thermal management, J. Spacecraft and Rockets 43 (2006).

[22] C.H. Li, G.P. Peterson, Experimental investigation of temperature and volume fraction variations on the effective thermal conductivity of nanoparticle suspensions (nanofluids), J. Appl. Phys. 99 (2006) 084314.

[23] D.-Hwang, K.S. Hong, H.-S. Yang, Study of thermal conductivity of nanofluids for the application of heat transfer fluids, Thermochimica Acta 455 (2007) 66-69.

[24] J.C. Maxwell, A Treatise on electricity and magnetism, II ed., Claredon, Oxford, 1881.

[25] C.-W. Nan, R. Birringer, D.R. Clarke, H. Gleiter, Effective thermal conductivity of particulate composites with interfacial thermal resistance, J. of Appl. Phys. 81 (1997) 6692-6699.

[26] Y. Benveniste, Effective thermal conductivity of composites with a thermal contact resistance between the constituents: Nondilute case, J. Appl. Phys. 61 (1987) 2840.

[27] S.H. Kim, S.R. Choi, D. Kim, Thermal Conductivity of Metal-Oxide Nanofluids: Particle Size Dependence and Effect of Laser Irradiation, J. Heat Transfer 129 (2007) 298-307.

[28] K.S. Hong, T.-K. Hong, H.-S. Yang, Thermal conductivity of Fe nanofluids depending on the cluster size of nanoparticles Appl. Phys. Lett. 88 (2006) 031901.

[29] S.K. Das, N. Putra, P. Thiesen, W. Roetzel, Temperature Dependence of Thermal Conductivity Enhancement for Nanofluids J. Heat Transfer 125 (2003) 567-574.

[30] A.G. Every, Y. Tzou, D.P.H. Hasselman, R. Raj, The effect of partcile size on the thermal conductivity of ZnS/diamond composites, Acta Metall. Mater. 40 (1992) 123-129.

[31] N.M. Sofian, M. Rusu, R. Neagu, E. Neagu, Metal powder-filled polyethylene composites. V. thermal

properties, J. Thermoplastic composite materials 14 (2001) 20-33.
[32] D.P.H. Hasselman, K.Y. Donaldson, Role of size in the effective thermal conductivity of composites with an interfacial thermal barrier, J. Wide Bandgap Mater. 7 (2000) 306-318.
[33] A.L. Geiger, D.P.H. Hasselman, K.Y. Donaldson, Effect of reinforcement particle size on the thermal conductivity of a particulate silicon-carbide reinforced aluminium-matrix composite, J. Mater. Sci. Lett. 12 (1993) 420-423.
[34] R. Pal, New models for thermal conductivity of particulate composites, J. Reinforced Plastics and Composites 26 (2007) 643-651.
[35] H. Zhang, X. Ge, H. Ye, Effectiveness of the heat conduction reinforcement of particle filled composites, Modelling Simul. Mater. Sci. Eng. 13 (2007) 401-412.
[36] D.H. Kumar, H.E. Patel, V.R.R. Kumar, T. Sundararajan, T. Pradeep, S.K. Das, Model for Heat Conduction in Nanofluids, Phys. Rev. Lett. 93 (2004) 144301.
[37] P. Bhattacharya, S.K. Saha, A. Yadav, P.E. Phelan, R.S. Prasher, Brownian dynamics simulation to determine the effective thermal conductivity of nanofluids, J. Appl. Phys 95 (2004) 6492.
[38] J. Koo, C. Kleinstreuer, A new thermal conductivity model for nanofluids J. Nanoparticle Res. 6 (2004) 577-588.
[39] S.P. Jang, S.U.S. Choi, Role of Brownian motion in the enhanced thermal conductivity of nanofluids, Appl. Phys. Lett. 84 (2004) 4316-4318.
[40] S.P. Jang, S.U.S. Choi, Effects of various parameters on nanofluid thermal conductivity, J. Heat Transfer 129 (2007) 617-623.
[41] R. Prasher, P. Bhattacharya, P.E. Phelan, Thermal Conductivity of Nanoscale Colloidal Solutions (Nanofluids), Phys. Rev. Lett. 94 (2005) 025901.
[42] R. Prasher, P. Bhattacharya, P.E. Phelan, Brownian-motion-based convective-conductive model for the effective thermal conductivity of nanofluids J. Heat Transfer 128 (2006) 588.
[43] C.H. Li, G.P. Peterson, Mixing effect on the enhacement of the effective thermal conductivity of nanoparticle suspensions (nanofluids), Int. J. Heat and Mass Transfer To appear (2007).
[44] H.E. Patel, T. Sundararajan, T. Pradeep, A. Dasgupta, N. Dasgupta, S.K. Das, A micro-convection model for thermal conductivity of nanofluids, Pramana - J. Phys. 65 (2005) 863.
[45] W. Yu, S.U.S. Choi, The role of interfacial layers in the enhanced thermal conductivity of nanofluids: A renovated Maxwell model, J. Nanoparticle Res. 5 (2003) 167-171.
[46] Q.-Z. Xue, Model for effective thermal conductivity of nanofluids, Phys. Lett. A 307 (2003) 313-317.
[47] H. Xie, M. Fujii, X. Zhang, Effect of interfacial nanolayer on the effective thermal conductivity of nanoparticle-fluid mixture, Int. J. Heat and Mass Transfer 48 (2005) 2926-2932.
[48] Q. Xue, W.-M. Xu, A model of thermal conductivity of nanofluids with interfacial shells, Materials Chem. and Phys. 90 (2005) 298-301.
[49] P. Tillman, J.M. Hill, Determination of nanolayer thickness for a nanofluid, Int. Comm. Heat and Mass Transfer 34 (2007) 399-407.
[50] J. Avsec, M. Oblak, The calculation of thermal conductivity, viscosity and thermodynamic properties for nanofluids on the basis of statistical nanomechanics, Int. J. Heat and Mass Transfer 50 (2007) 4331-4341.
[51] L. Gao, X.F. Zhou, Differential effective medium theory for thermal conductvity in nanofluids, Phys. Lett. A 348 (2006) 355-360.
[52] X.F. Zhou, L. Gao, Effective thermal conductivity in nanofluids of nonsperical particles with interfacial thermal resistance: Differential effective medium theory, J. Appl. Phys 1000 (2006) 024913.
[53] R. Prasher, W. Evans, P. Meakin, J. Fish, P. Phelan, P. Keblinski, Effect of aggregation on thermal conduction in colloidal nanofluids Appl. Phys. Lett. 89 (2006) 143119.
[54] J. Xu, B.-M. Yu, M.-J. Yun, Effect of clusters on thermal conductivity in nanofluids, Chin. Phys. Lett. 23 (2006) 2819-2822.
[55] Y. Feng, B. Yu, P. Xu, M. Zou, The effective thermal conductivity of nanofluids based on the nanolayer and the aggregation of nanopartciles, J. Phys. D: Appl. Phys. 40 (2007).
[56] J. Xu, B. Yu, M. Zou, P. Xu, A new model for heat conduction of nanofluids based on fractal distributions of nanoparticles, J. Phys. D: Appl. Phys. 39 (2006) 4486-4490.
[57] Y. Ren, H. Xie, A. Cai, Effective thermal conductivity of nanofluids containing spherical nanoparticles, J. Phys. D: Appl. Phys. 38 (2005) 3958-3961.
[58] B.-X. Wang, L.-P. Zhou, Z.-F. Peng, A fractal model for predicting the effective thermal conductivity of liquid with suspension of nanoparticles, Int. J. Heat and Mass Transfer 46 (2003) 2665-2672.
[59] Y. Xuan, Q. Li, W. Hu, Aggregation structure and thermal conductivity of nanofluids, AIChE J. 49 (2004) 1038-1043.
[60] Z. Hashin, S. Shtrikman, A variational approach to the theory of the effective magnetic permeability of multiphase materials, J. Appl. Phys 33 (1962) 3125.
[61] S.R. deGroot, P. Mazur, Nonequilibrium thermodynamics, Dover Publications, Inc., New York, 1984.
[62] A.L. DeVera, W. Strieder, Upper and lower bounds on the thermal conductvity of a random, two-phase material, J. Phys. Chem. 81 (1977) 1783.
[63] A. Griesinger, W. Hurler, M. Pietralla, A photothermal method with step heating for measuring the

thermal diffusivity of anisotropic solids, Int. J. Heat and Mass Transfer 40 (1997) 3049-3058.
[64] J.K. Carson, S.J. Lovatt, D.J. Tanner, A.C. Cleland, Thermal conductivity bounds for isotropic porous materials, Int. J. Heat and Mass Transfer 48 (2005) 2150-2158.
[65] S. Torquato, M.D. Rintoul, Effect of interface on the properties of composite media, Phys. Rev. Lett. 75 (1995) 4067-4070.
[66] K. Raghavan, K. Foster, K. Motakabbir, M. Berkowitz, Structure and dynamics of water at the Pt(111) interface: Molecular dynamics study, J. Chem. Phys. 94 (1991) 2110.
[67] M.F. Reedijk, J. Arsic, F.F.A. Hollander, S.A.d. Vries, E. Vlieg, Liquid order at the interface of KDP crystals with water: evidence for icelike layers, Phys. Rev. Lett. 90 (2003) 066103.
[68] H. Mo, G. Evmenenko, P. Dutta, Ordering of liquid squalane near a solid surface, Chem. Phys. Letters 415 (2005) 106-109.
[69] C.-J. Yu, A.G. Richter, J. Kmetko, S.W. Dugan, A. Datta, P. Dutta, Structure of interfacial liquids: X-ray scattering studies, Phys. Rev. E 63 (2001) 021205.
[70] J. Eapen, J. Li, S. Yip, Beyond Maxwell limit: Thermal conduction in nanofluids with percolating fluid structures, arXiv:0707.2164v1 (2007).
[71] L. Xue, P. Keblinski, S.R. Phillpot, S.U.S. Choi, J.A. Eastman, Effect of liquid layering at the liquid–solid interface on thermal transport, Int. J. Heat and Mass Transfer 47 (2004) 4277-4283.
[72] W. Evans, J. Fish, P.Keblinski, Thermal conductivity of ordered molecular water J. Chem. Phys. 126 (2007) 154504.
[73] H.Mo, G. Evmenenko, S. Kewalramani, K. Kim, S.N. Ehrlich, P. Dutta, Observation of Surface Layering in a Nonmetallic Liquid Phys. Rev. Lett. 96 (2006) 096107.
[74] X.-Q. Wang, A.S. Mujumdar, Heat transfer characteristics of nanofluids: a review Int. J. Therm. Sci. In Press (2006).
[75] P. Keblinski, D.G. Cahill, Comment on "Model for Heat Conduction in Nanofluids", Phys. Rev. Lett. 95 (2005) 209401.
[76] S. Bastea, Comment on "Model for Heat Conduction in Nanofluids", Phys. Rev. Lett. 95 (2005) 019401.
[77] M.H. Ernst, R. Brito, New Green-Kubo formulas for transport coefficients in hard-sphere, Langevin fluids and the likes, Europhys. Lett. 73 (2006) 183-189.
[78] C. Marsh, G. Backx, M. Ernst, Fokker-Planck-Boltzmann equation for dissipative particle dynamics, Europhys. Lett. 38 (1997) 411.
[79] K.I. Morozov, On the theory of the Soret effect in colloids, Springer-Verlag, Berlin, 2002.
[80] R. Piazza, 'Thermal forces': colloids in temperature gradients, J. Phys.: Condens. Matter 16 (2004) S4195.
[81] W.F.S. Kenneth C. Hass, Alessandro Curioni, Wanda Andreoni The Chemistry of Water on Alumina Surfaces: Reaction Dynamics from First Principles, Science 282 (1998) 265-268.
[82] W. Evans, J. Fish, P. Keblinski, Role of Brownian motion hydrodynamics on nanofluid thermal conductivity, Appl. Phys. Lett. 88 (2006) 093116.
[83] P. Keblinski, S.R. Phillpot, S.U.S. Choi, J.A. Eastman, Mechanisms of heat flow in suspensions of nano-sized particles (nanofluids) Int. J. Heat and Mass Transfer 45 (2002) 855-863.
[84] J. Eapen, W.C. Williams, J. Buongiorno, L.-W. Hu, S. Yip, R. Rusconi, R. Piazza, Mean-field versus microconvection effects in nanofluid thermal conduction, Phys. Rev. Lett. 99 (2007) 095901.
[85] P. Keblinski, J. Thomin, Hydrodynamic field around a Brownian particle, Phys. Rev. E 73 (2006) 010502(R).
[86] J. Buongiorno, Convective transport in nanofluids, J. Heat Transfer 128 (2006) 240.
[87] I. Goldhirsch, D. Ronis, Theory of thermophoresis. I. General considerations and mode-coupling analysis, Phys. Rev. A 27 (1983) 1616-1634.
[88] F. Lin, G.S. Bhatia, J.D. Ford, Thermal conductivities of powder-filled epoxy resins, J. Appl. Polymer. Sci. 49 (1993) 1901-1908.
[89] N.M. Sofian, M. Rusu, R. Neagu, E. Neagu, Metal powder-filled polyethylene composites. V. thermal properties, Journal of Thermoplastic Composite Materials 14 (2001) 20-33.
[90] A. Boudenne, L. Ibos, M. Fois, E. Gehin, J.-C. Majeste, Thermophysical properties of polypropylene/aluminum composites, J. Polymer Sci. Part B. 42 (2004) 722-732.
[91] H.S. Tekce, D. Kumlutas, I.H. Tavman, Effect of particle shape on thermal conductivity of copper reinforced polymer composites, Journal of Reinforced Plastics and Composites 26 (2007) 113-121.
[92] Y. Xu, D.D.L. Chung, C. Mroz, Thermally conducting aluminum nitride polymer-matrix composites, Composites: Part A 32 (2001) 1749-1757.
[93] Y. Agari, T. Uno, Estimation of thermal conductivities of filled polymers, J. Appl. Polymer. Sci. 32 (1986) 5705-5712.
[94] T.G. Cowling, P. Gray, P.G. Wright, The physical significance of formulae for the thermal conductivity and viscosity of gaseous mixtures, Proc. Royal Soc. London, Series A 276 (1963) 69-82.
[95] J.D. Pandey, R.K. Mishra, Theoretical evaluation of thermal conductivity and diffusion coefficient of binary liquid mixtures, Phys. and Chem. of Liquids 43 (2005) 49-57.
[96] M.J. Assael, E. Charitidou, W.A. Wakeham, Absolute measurements of the thermal conductivity of mixtures of alcohols with water, Int. J. Thermophysics 10 (1989) 793-803.
[97] C.C. Li, Thermal conductivity of liquid mixtures, AIChE J. 22 (1976) 927-930.

[98] K. Jagannadham, H. Wang, Thermal resistance of interfaces in AlN–diamond thin film composites, J. Appl. Phys 91 (2002).
[99] Z. Ge, D.G. Cahill, P.V. Braun, Thermal conductance of hydrophilic and hydrophobic interfaces, Phys. Rev. Lett. 96 (2006) 186101.
[100] M.B. Bryning, D.E. Milkie, M.F. Islam, J.M. Kikkawa, A.G. Yodh, Thermal conductivity and interfacial resistance in single-wall carbon nanotube epoxy composites, Appl. Phys. Lett. 87 (2005) 161909.
[101] S.T. Huxtable, D.G. Cahill, S. Shenogin, L. Xue, R. Ozisik, P. Barone, M. Usrey, M.S. Strano, G. Siddons, M. Shim, P. Keblinski, Interfacial heat flow in carbon nanotube suspension, Nature Materials 2 (2003).
[102] C.-W. Nan, G. Liu, Y. Lin, M. Li, Interface effect on thermal conductivity of carbon nanotube composites, Appl. Phys. Lett 85 (2004) 3549-3551.
[103] H. Masuda, A. Ebata, K. Teramae, N. Hishinuma, Alteration of thermal conductivity and viscosity of liquid by dispersing ultra-fine particles (Dispersion of $\gamma$-Al2O3 , SiO2 , and TiO2 ultra-fine particles), Netsu Bussei (Japan) 7 (1993) 227-233.
[104] S. Lee, S.U.S. Choi, S. Li, J.A. Eastman, Measuring thermal conductivity of fluids containing oxide nanoparticles, J. Heat Transfer 121 (1999) 280–289.
[105] J.A. Eastman, S.U.S. Choi, S. Li, L.J. Thompson, S. Lee, Enhanced thermal conductivity through the development of nanofluids, Materials Research Society (MRS): Fall Meeting, Boston, USA, 1996, pp. 3–11.
[106] D. Wen, Y. Ding, Experimental investigation into convective heat transfer of nanofluids at the entrance region under laminar flow conditions, Int. J. Heat and Mass Transfer 47 (2004) 5181-5188.
[107] D. Wen, Y. Ding, Natural convective heat transfer of suspensions of titanium dioxide nanoparticles (nanofluids), IEEE Trans. Nanotech. 5 (2006) 220-227.
[108] S. Shaikh, K. Lafdi, R. Ponnappan, Thermal conductivity improvement in carbon nanoparticle doped PAO oil: An experimental study, J. Appl. Phys. 101 (2007) 064302.
[109] S.U.S. Choi, Z.G. Zhang, W. Yu, F.E. Lockwood, E.A. Grulke, Anomalous thermal conductivity enhancement in nanotube suspensions, Appl. Phys. Lett. 79 (2001) 2252-2254.
[110] Y. Hwang, J.K. Lee, C.H. Lee, Y.M. Jung, S.I. Cheong, C.G. Lee, B.C. Ku, S.P. Jang, Stability and thermal conductivity characteristics of nanofluids, Thermochimica Acta 455 (2007) 70-74.
[111] D. Wen, Y. Ding, Effective thermal conductivity of aqueous suspensions of carbon nanotubes (carbon nanotube nanofluids), J. Thermophys. Heat Transfer 18 (2004) 481-485.
[112] D.A. McQuarrie, Statistical Mechanics University Science Books, California, 2000.
[113] D.C. Rapaport, The art of molecular dynamics simulation, 2 ed., Cambridge University Press, Cambridge, 2004.
[114] H. Kaburaki, J. Li, S. Yip, H. Kimizuka, Dynamical thermal conductivity of Argon crystal, J. Appl. Phys. (to appear) (2007).
[115] J. Eapen, J. Li, S. Yip, Mechanism of Thermal Transport in Dilute Nanocolloids Phys. Rev. Lett. 98 (2007) 028302.
[116] J. Li, L.J. Porter, S. Yip, Atomistic modeling of finite-temperature properties of crystalline beta-SiC, II. thermal conductivity and effects of point defects, J. of Nucl. Mater. 255 (1998) 139-152.
[117] M. Vladkov, J.-L. Barrat, Modeling transient absorption and thermal conductivity in a simple nanofluid, Nano Lett. 6 (2006) 1224-1228.
[118] W.C. Williams, Experimental and theoretical investigations of transport phenomena in nanoparticle colloids (nanofluids), PhD Thesis, Department of Nuclear Science and Engineering, Massachusetts Institute of Technology, Cambridge, 2006.
[119] D. Bertolini, A. Tani, Thermal conductivity of water: Molecular dynamics and generalized hydrodynamics results, Phys. Rev. E 56 (1997) 4135.
[120] R. Prasher, P.E. Phelan, P. Bhattacharya, Effect of aggregation kinetics on the thermal conductivity of nanoscale colloidal solutions (nanofluid), Nano Lett. 6 (2006) 1529 - 1534.
[121] L. Li, D.D.L. Chung, Thermally conducting polymer-matrix composites containing both AlN particles and SIC whiskers, Journal of Electronic Materials 23 (1994) 557-564.
[122] National Institute of Standards and Technology (NIST): fluid properties, http://webbook.nist.gov/chemistry/fluid/ (2007).
[123] CRC Handbook of Chemistry & Physics, http://www.hbcpnetbase.com (2007).
[124] The A-Z of materials, http://www.azom.com (2007).
[125] K. Kwak, C. Kim, Viscosity and thermal conductivity of copper oxide nanofluid dispersed in ethylene glycol, Korea-Australian Rheology Journal 17 (2005) 35-40.
[126] R.C. Yu, N. Tea, M.B. Salamon, D. Lorents, R. Malhotra, Thermal conductivity of single crystal C60, Phys. Rev. Lett. 68 (1992) 2050-2053.

# Appendix

Table 1. Thermal conductivity data for solid composites

| Material | κ (W/m-K) | Reference |
|---|---|---|
| Aluminum | 237 | [90] |
| Alumina | 33.095 | [93] |
| Graphite | 209.2 | [93] |
| Cupric oxide | 9.21 | [88] |
| Copper | 384 | [91] |
| Aluminum nitride | 200 | [121] |
| Diamond (polycrystal) | 600 | [30] |
| Zn | 116 | [89] |
| Polypropylene | 0.239 | [90] |
| Polyethylene | 0.505/0.291 | [89, 93] |
| Epoxy | 0.221 | [88] |
| Polyamide | 0.32 | [91] |
| Polyvinylidene fluoride | 0.12 | [92] |
| Zinc sulphide | 17.065 | [30] |

Table 2. Thermal conductivity data for liquid mixtures

| Material | κ (W/m-K) | Reference |
|---|---|---|
| Water | 0.61 | 25 ºC, [122] |
| Ethanol | 0.161 | 32.6 ºC, [96] |
| Acetone | 0.154 | 40 ºC, [97] |
| Methanol | 0.198 | 28.1 ºC, [96] |
| Propylene glycol | 0.199 | 40 ºC, [97] |
| Ethylene glycol | 0.252 | 40 ºC, [97] |
| Propan-2-ol | 0.137 | 40 ºC, [97] |
| Formamide | 0.352 | 40 ºC, [97] |

Table 3. Thermal conductivity data for nanofluids

| Material | κ (W/m-K) | Reference |
|---|---|---|
| Water | 0.61 | 25 ºC, [122] |
| Ethylene glycol (EG) | 0.25 | 25 ºC (Measured data) |
| Oil | 0.11-0.15 | [29, 108-110] |
| $Al_2O_3$ | 36 | Representative value, [93, 123, 124] |
| CuO | 20 | Representative value, [29, 125] |
| $ZrO_2$ | 2 | Representative value, [124] |
| $TiO_2$ | 11.7 | Representative value, [124] |
| $SiO_2$ | 1.7 | Representative value, [84, 110] |
| $Fe_3O_4$ | 7.0 | [123] |
| Au | 317 | [123] |
| Ag | 429 | [123] |
| Al | 237 | [123] |
| Fe | 80.2 | [123] |
| Carbon nanotubes (CNT) | 2000 | Representative value, [109] |
| Diamond | 600 | Polycrstalline, [30] |
| Fullerenes ($C_{60}$) | 0.4 | [126] |
| MFA | 0.2 | [84] |
| $Al_2Cu$ | 237 | Assumed value of Al, [123] |